\documentclass[iop,numberedappendix,appendixfloats,twocolappendix]{emulateapj}
\usepackage{graphicx,subfigure,amsmath, amsfonts, amssymb,aas_macros,times,natbib}
\usepackage[export]{adjustbox}
\usepackage[usenames,dvipsnames]{xcolor}
\usepackage[toc,page]{appendix}
\usepackage[bookmarks=false]{hyperref}
\hypersetup{
  colorlinks= true, 
  urlcolor  = Blue, 
  filecolor=black,
  runcolor=blue,
  linkcolor = Maroon, 
  citecolor = BlueViolet 
} 

\newcommand{\nii}{\hbox{[N\,\sc{ii}]}}
\newcommand{\sii}{\hbox{[S\,\sc{ii}]}}
\newcommand{\oiii}{\hbox{[O\,\sc{iii}]}}
\newcommand{\oi}{\hbox{[O\,\sc{i}]}}
\newcommand{\ha}{H$\alpha$}
\newcommand{\hb}{H$\beta$}
\newcommand{\nad}{Na\,{\sc i}\,D}
\newcommand{\ewhd}{\hbox{EW(H$\delta$)\,$>$\,5\AA}}
\newcommand{\lcomhtwo}{$L_{\rm CO}$--$M_{\rm H_2}$}

\shorttitle{SPOGS II: CO(1--0) Observations of SPOGs}
\shortauthors{Alatalo et al.}

\begin{document}


\title{Shocked POststarburst Galaxy Survey II: the Molecular Gas Content and Properties of a Subset of SPOGs}

\author{Katherine Alatalo,$^{1}$\altaffilmark{$\dagger$} Ute Lisenfeld,$^{2}$ Lauranne Lanz,$^{3}$ Philip~N. Appleton,$^{3}$ Felipe Ardila,$^{4}$ Sabrina L. Cales,$^{5}$  Lisa~J. Kewley,$^{6}$ Mark Lacy,$^{7}$  Anne~M. Medling,$^{6}$ Kristina Nyland,$^{7}$ Jeffrey~A. Rich,$^{1,3}$ \& C. Meg Urry$^{5}$}

\affil{
$^{1}$Observatories of the Carnegie Institution of Washington, 813 Santa Barbara Street, Pasadena, CA 91101, USA\\
$^{2}$Departamento de F\'isica Te\'orica y del Cosmos, Universidad de Granada, Granada, Spain\\
$^{3}$Infrared Processing and Analysis Center, California Institute of Technology, Pasadena, California 91125, USA\\
$^{4}$Department of Astrophysical Sciences, Princeton University, Peyton Hall, 4 Ivy Lane, Princeton, NJ 08544, USA\\
$^{5}$Department of Astronomy, Yale University, New Haven, CT 06511 USA\\
$^{6}$Research School of Astronomy and Astrophysics, Australian National University, Cotter Rd., Weston ACT 2611, Australia\\
$^{7}$National Radio Astronomy Observatory, 520 Edgemont Road, Charlottesville, VA 22903, USA}

\altaffiltext{$\dagger$}{Hubble fellow}
\email{kalatalo@carnegiescience.edu}
\slugcomment{Accepted by the Astrophysical Journal, June 2, 2016}

\begin{abstract}
We present CO(1--0) observations of objects within the Shocked POststarburst Galaxy Survey taken with the Institut de Radioastronomie Millim\'etrique (IRAM) 30m single dish and the Combined Array for Research for Millimeter Astronomy (CARMA) interferometer. Shocked Poststarburst Galaxies (SPOGs) represent a transitioning population of galaxies, with deep Balmer absorption (EW$_{\rm H\delta}$\,$>$\,5\AA), consistent with an intermediate-age (A-star) stellar population, and ionized gas line ratios inconsistent with pure star formation. The CO(1--0) subsample was selected from SPOGs detected by the {\em Wide-field Infrared Survey Explorer} with 22$\mu$m flux detected at a signal-to-noise (S/N) $>$3. Of the 52 objects observed in CO(1--0), 47 are detected with S/N$>$3. A large fraction (37--46$\pm$7\%) of our CO-SPOG sample were visually classified as morphologically disrupted. The H$_2$ masses detected were between 10$^{8.7-10.8}$~$M_\odot$, consistent with the gas masses found in normal galaxies, though approximately an order of magnitude larger than the range seen in poststarburst galaxies.  When comparing the 22$\mu$m and CO(1--0) fluxes, SPOGs diverge from the normal star-forming relation, having 22$\mu$m fluxes in excess of the relation by a factor of $\langle\epsilon_{\rm MIR}\rangle$\,=\,4.91$^{+0.42}_{-0.39}$, suggestive of the presence of active galactic nuclei (AGN). The \nad\ characteristics of CO-SPOGs show that it is likely that many of these objects host interstellar winds. Objects with the large \nad\ enhancements also tend to emit in the radio, suggesting possible AGN-driving of neutral winds.
\end{abstract}




\section{Introduction}
The bimodality of morphological classifications of galaxies has long been known \citep{hubble26}.  Typical galaxies are either classified as ``late-type'' galaxies, or ``early-type'' galaxies. ``Late-types'' have thin disks and exhibit spiral structure and blue colors. ``Early-types'' tend to be more ellipsoidal, contain smoother isophotes, and exhibit redder colors. Galaxies also bifurcate across colors with a red and blue population \citep{baade58,holmberg58,tinsley78,strateva+01,baldry+04,faber+07} based primarily on their star formation properties, and few galaxies have intermediate, ``green valley'' colors \citep{bell+03}. The morphological and color bimodalities seem to indicate that galaxies transform rapidly between the blue cloud and red sequence \citep{martin+07}. Star-forming galaxies are blue in color and span a large range of magnitudes, known as the ``blue cloud''.  Red sequence galaxies, on the other hand, inhabit a well defined region with much smaller variation in both color and magnitude.  As in the case of the morphological classification of galaxies, the color bimodality seen in galaxies can be explained simply by quenching star formation.  Once a star-forming galaxy has had its star formation quenched, it quickly migrates from the blue cloud and becomes a red sequence galaxy \citep{harker+06}.  The morphological and color properties of individual galaxies are usually well-matched, with early-types also being red sequence galaxies, and late-types also being star-forming galaxies.

\begin{figure*}[t]
\centering
\includegraphics[width=\textwidth,clip,trim=0cm 0.2cm 6.9cm 0.1cm]{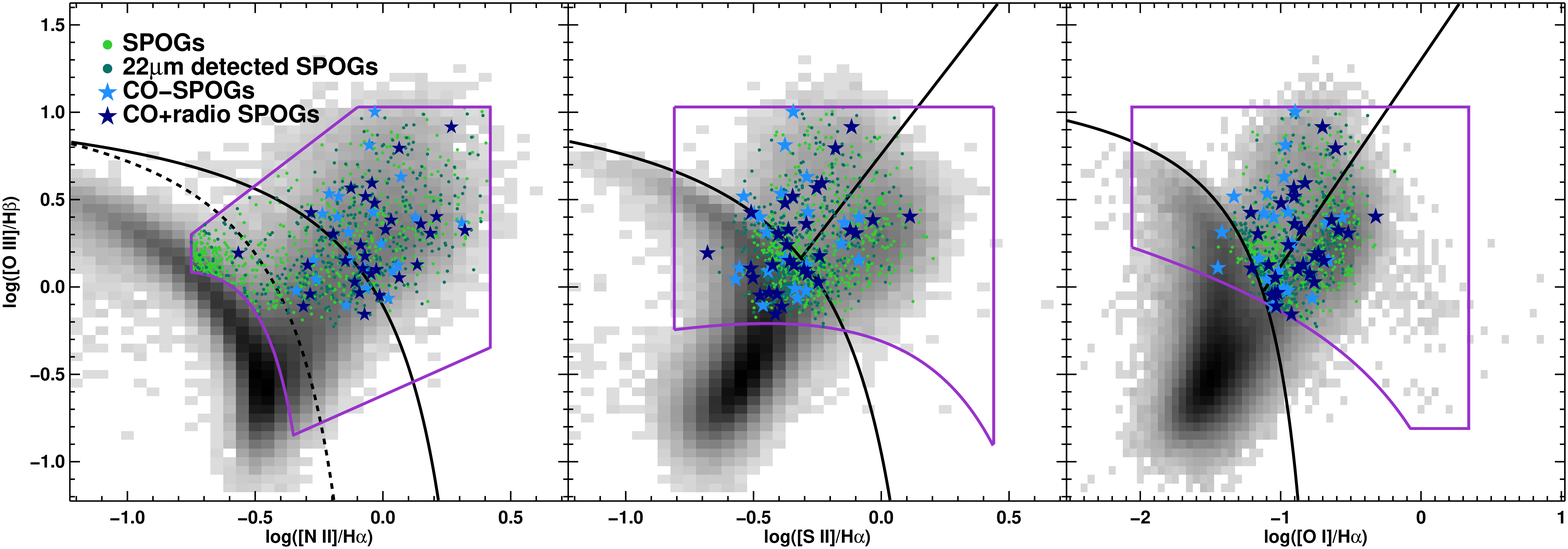}
\caption{The ionized gas line ratios of the ELG (grayscale) and SPOG samples (green dots \citealt{a16_sample}), including \nii/\ha\ vs. \oiii/\hb\ (left; \citealt{bpt}), \sii/\ha\ vs. \oiii/\hb\ (middle), and \oi/\ha\ vs. \oiii/\hb\ (right; \citealt{veilleux+87}), overlaid with the line diagnostic models of \citet{kauffmann+03,kewley+06}. The purple line defines the boundaries of the shock models, SPOG criterion \citep{allen+08,rich+11,a16_sample}. {\em WISE} 22$\mu$m-detected SPOGs are shown in dark green. CO-SPOGs (light blue stars), and 1.4\,GHz radio-matched CO-SPOGs (dark blue stars) are also shown. CO-SPOGs span the ionized gas diagnostic space of the larger SPOG sample. There is also little difference between the radio matched and unmatched 22$\mu$m-detected SPOGs.}
\label{fig:wise_bpts}
\end{figure*}

Many transformational paths have been proposed, including a merger between two late-type galaxies into an early-type in simulation \citep{toomre72,springel+05}, ram pressure stripping due to falling into a galaxy cluster \citep{bekki+02,park+07,blanton+moustakas09,chung+09b,kenney+14}, morphological quenching \citep{martig+09,martig+13}, tidal disruption and harassment through group interactions \citep{zabludoff+mulchaey98,hickson+92,rasmussen+08,bitsakis+10,bitsakis+14}, and Active Galactic Nucleus (AGN) feedback \citep{fischer+10,feruglio+10,feruglio+15,sturm+11,alatalo+11,aalto+12,aalto_1377,cicone+12,cicone+14,alatalo15}. In the modern universe ({\em z}\,$\sim$\,0), this transformation appears to be one-way \citep{appleton+14,young+14}. Thus, it is essential to understand all pathways that can lead a blue late-type galaxy to become a red early-type.

\begin{figure}[b]
\centering
\includegraphics[width=0.49\textwidth,clip,trim=0cm 0cm 0cm 0cm]{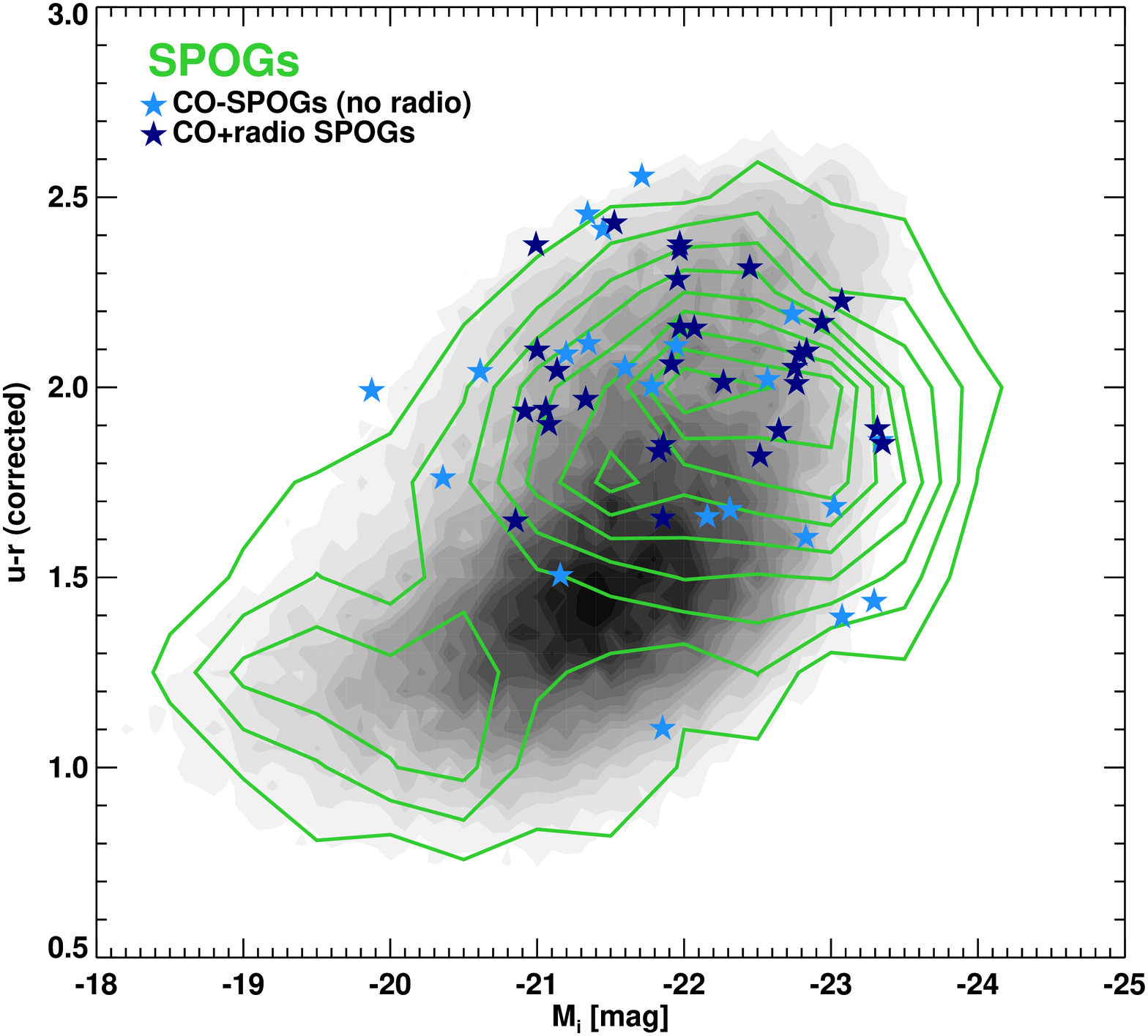}
\caption{Color-magnitude diagram from the parent emission line galaxy sample (grayscale; \citealt{a16_sample}) with the distribution of SPOGs overlaid (green contours), with $M_i$ representing the (uncorrected) absolute i-band magnitude. The CO-SPOGs are overlaid and color-coded based on their radio detections, with FIRST-detected SPOGs shown as dark blue stars, and radio non-detected CO SPOGs as light blue stars. The CO-SPOGs tend to be more massive than SPOGs in general, but not significantly, and their colors trace those colors of the underlying SPOGs distribution fairly well.}
\label{fig:co_cmd}
\end{figure}

\begin{figure*}
\centering
\includegraphics[width=\textwidth]{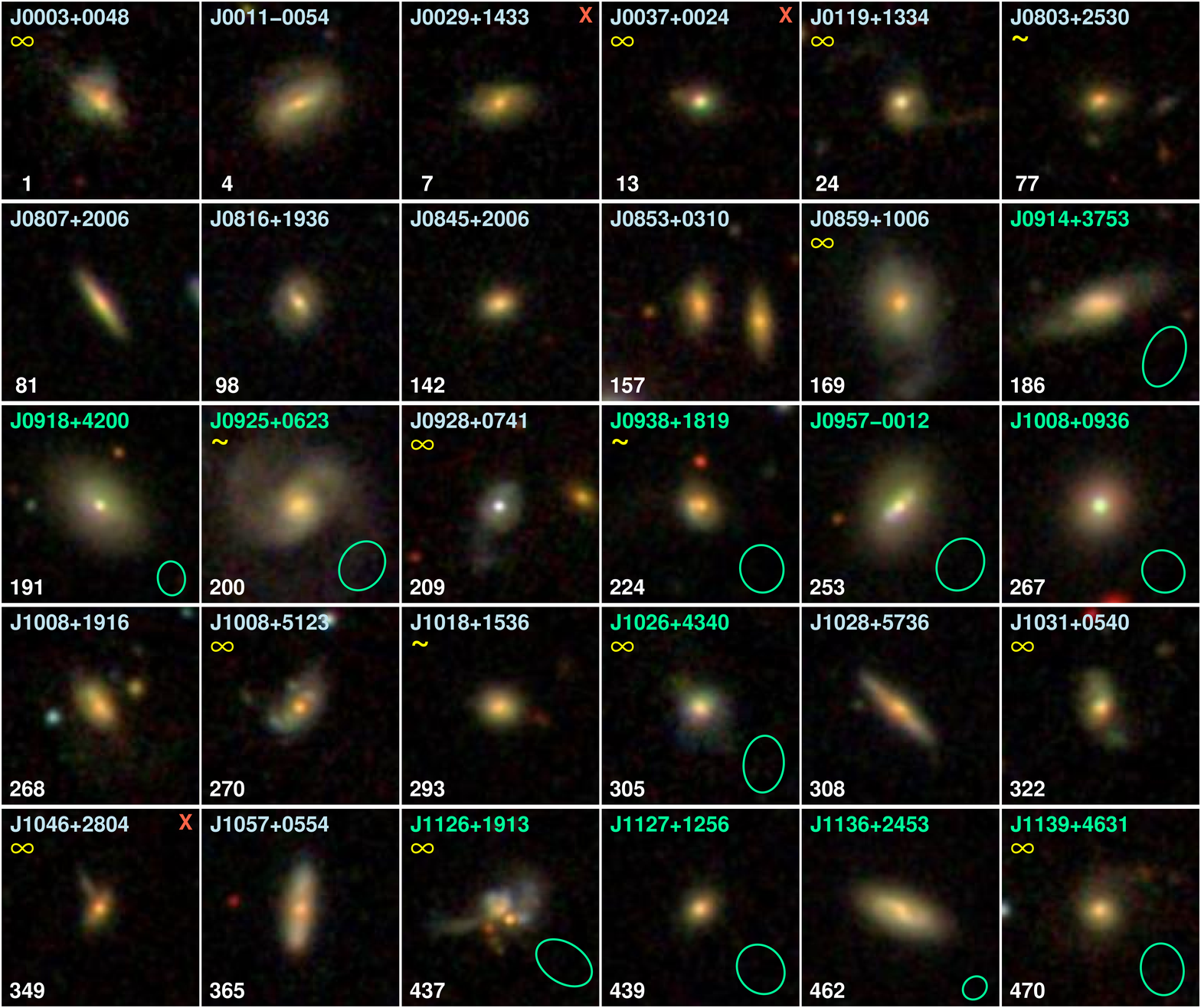}
\caption{SDSS {\em g\,r\,i} 3-color thumbnails of the 52 CO-SPOGs. The field of view of all thumbnails is 30$''$. Objects detected by CARMA are labeled in green, with the CARMA beam shown as an ellipse in the bottom right. Objects observed with the IRAM~30m are labeled in pale blue. The SPOG index from \citet{a16_sample} are labeled in white (bottom left). Non-detections are demarcated with a small \textcolor{Red}{\bf X} in the top right. Objects that have been morphologically classified as clearly disrupted have a small yellow symbol (\textcolor{Dandelion}{$\mathbf{\infty}$}) below the object name. Those that are classified as possibly disrupted, a (\textcolor{Dandelion}{$\mathbf{\sim}$}). Many SPOGs show signs of interaction, including tidal tails, dust lanes, and morphological disruption. Most galaxies appear to be red in color, with many also showing signs of peaked, bright nuclei, possibly due to the presence of an AGN.}
\label{fig:thumbs}
\end{figure*}
\renewcommand{\thefigure}{\arabic{figure} (Cont)}
\addtocounter{figure}{-1}
\begin{figure*}
\centering
\includegraphics[width=\textwidth,clip,trim=0cm 10cm 0cm 0cm]{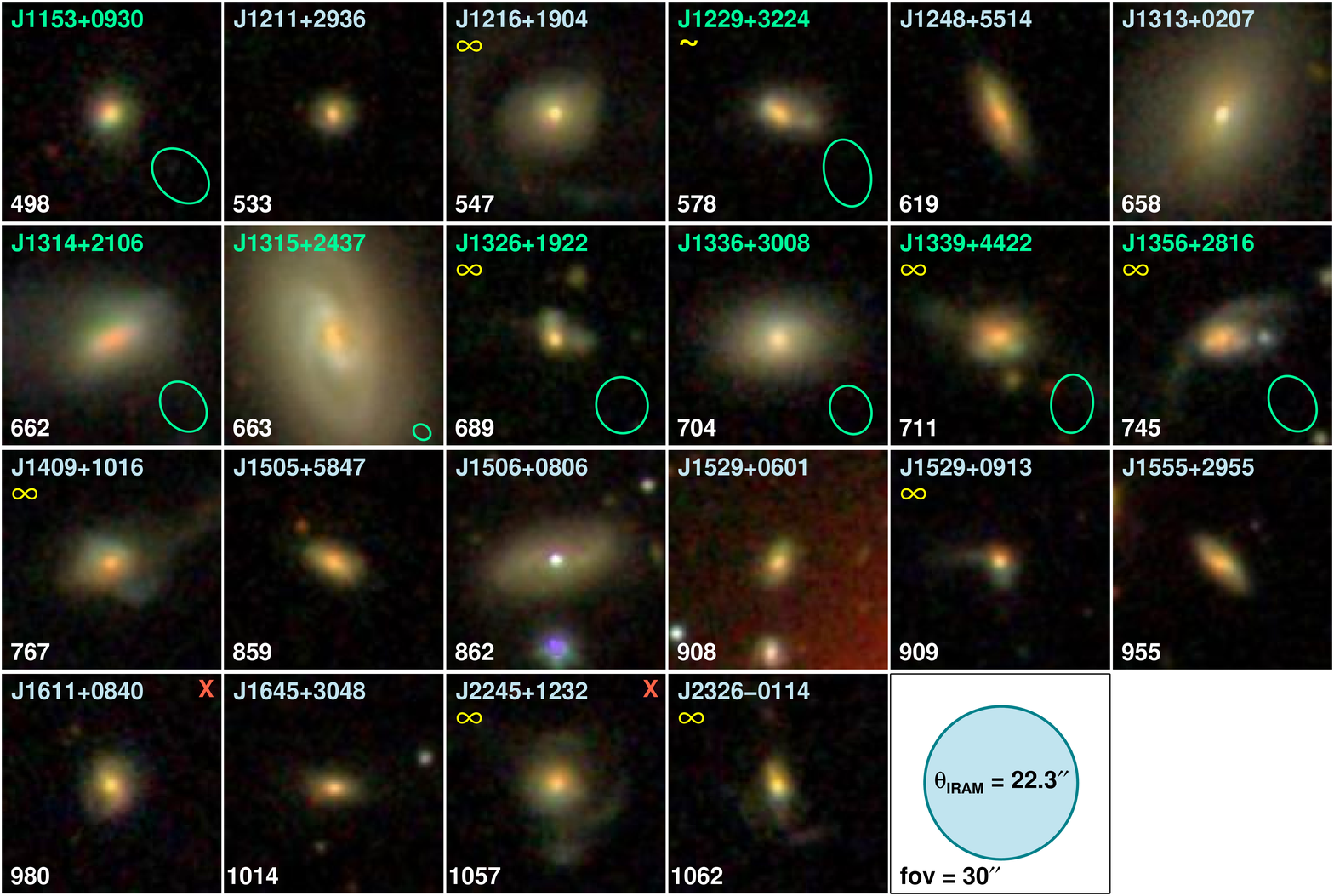}
\vskip -1mm
\caption{Continuation of the SDSS {\em g\,r\,i} 3-color images of CO-SPOGs. The last panel shows the 22.3$''$ single dish primary beam at $\nu_{\rm obs}$\,=\,110\,GHz for the SPOGs observed with the IRAM~30m (labeled in pale blue).}
\end{figure*}
\renewcommand{\thefigure}{\arabic{figure}}

\citet{schawinski+14} showed that the majority of galaxies with green colors were in fact normal spiral galaxies, with normal star formation rates that had built up a significant population of intermediate and older stars.  For this reason, optical color selection alone is not able to definitively identify a transitioning galaxy.  With the onset of the {\em Wide-field Infrared Survey Explorer} ({\em WISE}; \citealt{wise}) mission, evidence mounted that mid-IR colors could be used to identify transitioning galaxies. \citet{a14_irtz} presented existence of a prominent bifurcation between star-forming spiral galaxies and quiescent early-type galaxies in the {\em WISE} [4.6]--[12]$\mu$m bands, deeming this the ``Infrared Transition Zone'' (IRTZ).  Early-type galaxies that were within this IRTZ were found to have red optical colors (also described in \citealt{ko+13,yesuf+14}), indicating that galaxies must traverse  the optical green valley before the IRTZ.


Searches aiming to identify recently quenched galaxies have focused on objects with deep Balmer absorption lines, consistent with the presence of intermediate A-stars \citep{obafgkm} stellar population \citep{vazdekis+10}, and a lack of ionized gas emission lines usually associated with star formation (nebular lines such as \ha\ or [O\,{\sc ii}]$\lambda$3727; \citealt{dressler+gunn83,zabludoff+96,quintero+04,goto05,goto07}). While this selection is able to find recently quenched galaxies, it provides an incomplete picture of transforming objects.  Traditional ``E+A'' or ``K+A'' searches miss objects that exhibit line emission, including AGNs \citep{wild+09,kocevski+11,cales+11,cales+13}, emission from post-asymptotic branch (post-AGB) stars \citep{yan+06} and shocks \citep{allen+08,rich+11,a16_sample}. Traditional poststarburst searches also miss galaxies that have quenched but still have some low-level H$\alpha$ or [O\,{\sc ii}] emission \citep{yesuf+14,rowlands+15}.

While the most common picture of galaxy evolution involves the exhaustion of the star-forming fuel prior to the complete cessation of star formation \citep{hopkins+06}, observations of molecular gas in ``red and dead'' galaxies seems to indicate that a molecular reservoir can remain intact after galaxy transformation \citep{combes+07,young+11,davis+11}, even without re-accretion of new molecular material, although in these cases the molecular gas mass is no more than 1\% of the total stellar mass. More recently, \citet{french+15} and \citet{rowlands+15} have been able to show that significant reservoirs of molecular gas remain in post-transition objects (in these cases, poststarburst galaxies), calling into question the need to completely deplete (or significantly reduce) the molecular reservoir within a galaxy to cause star formation quenching and galaxy transition. 

This reservoir of gas could also be explained if poststarburst galaxies originate when early-type galaxies accrete material from the environment and go through a brief starburst phase, which was suggested by recent observations of a large sample of galaxies by \citet{dressler+13}. Recently, interferometric molecular gas observations have shown that the star formation within some quenched objects is suppressed, with inefficiencies of factors of 20--70 \citep{a15_hcgco,a15_sfsupp,guillard+15,aalto+16,salome+16,lanz+16}, leading to the possibility that star formation-suppressed molecular reservoirs are a common part of galaxy transformation. Although a larger sample of galaxies must be studied to determine whether this occurs only in rare and energetic objects.

\begin{table*}[t!]
\centering
\caption{CO-SPOG Sample Properties} \vspace{-1mm}
\begin{tabular}{l c c c c r r r r c}
\hline \hline
\# & SPOG & Telescope & RA & Dec & {\em z}~~~~~ & F$_{\rm 22\mu m}$ & F$_{\rm 1.4}$~ & log($M_{\star}$) & Morphology \\
& Name & & (J2000) & (J2000) & & (mJy) & (mJy) & ($M_\odot$) & \\
(1) & (2) & (3) & (4) & (5)~~~ & (6)~~ & (7)~~ & (8)~~ & (9) & (10) \\
\hline
1 & J0003+0048 & I & 00\,03\,18.21 & +00\,48\,44.2 & 0.139 & 14.2$\pm$1.6 & 3.71$\pm$0.096 & 10.72 & $$\checkmark$ $ \\
4 & J0011-0054 & I & 00\,11\,45.21 & -00\,54\,44.2 & 0.048 & 20.5$\pm$1.3 & 2.08$\pm$0.14 & 10.34 & $... $ \\
7 & J0029+1433 & I & 00\,29\,28.96 & +14\,33\,42.8 & 0.143 & 10.2$\pm$2.0 & ... & 10.69 & $... $ \\
13 & J0037+0024 & I & 00\,37\,07.82 & +00\,24\,36.3 & 0.081 & 21.1$\pm$1.5 & ... & 10.18 & $$\checkmark$ $ \\
24 & J0119+1334 & I & 01\,19\,56.76 & +13\,34\,31.4 & 0.191 & 3.4$\pm$0.3 & ... & 10.95 & $$\checkmark$ $ \\
77 & J0803+2530 & I & 08\,03\,59.61 & +25\,30\,51.4 & 0.135 & 11.3$\pm$1.2 & 1.13$\pm$0.14 & 10.93 & $? $ \\
81 & J0807+2006 & I & 08\,07\,24.45 & +20\,06\,08.2 & 0.066 & 21.2$\pm$1.3 & 3.48$\pm$0.14 & 10.39 & $... $ \\
98 & J0816+1936 & I & 08\,16\,03.14 & +19\,36\,43.2 & 0.113 & 12.3$\pm$0.99 & ... & 10.40 & $... $ \\
142 & J0845+2006 & I & 08\,45\,45.38 & +20\,06\,10.4 & 0.123 & 17.7$\pm$2.2 & ... & 10.61 & $... $ \\
157 & J0853+0310 & I & 08\,53\,56.80 & +03\,10\,33.6 & 0.129 & 3.7$\pm$0.2 & 1.25$\pm$0.15 & 10.88 & $... $ \\
169 & J0859+1006 & I & 08\,59\,42.62 & +10\,06\,43.5 & 0.055 & 82.5$\pm$2.3 & 2.73$\pm$0.14 & 10.54 & $$\checkmark$ $ \\
186 & J0914+3753 & C & 09\,14\,07.22 & +37\,53\,09.9 & 0.072 & 29.4$\pm$1.4 & 2.60$\pm$0.16 & 10.30 & $... $ \\
191 & J0918+4200 & C & 09\,18\,49.99 & +42\,00\,43.5 & 0.041 & 40.4$\pm$1.5 & ... & 10.30 & $... $ \\
200 & J0925+0623 & C & 09\,25\,18.31 & +06\,23\,34.0 & 0.076 & 27.0$\pm$1.3 & ... & 10.51 & $? $ \\
209 & J0928+0741 & I & 09\,28\,19.53 & +07\,41\,58.5 & 0.105 & 20.4$\pm$1.1 & ... & 10.11 & $$\checkmark$ $ \\
224 & J0938+1819 & C & 09\,38\,19.87 & +18\,19\,52.6 & 0.089 & 5.3$\pm$0.2 & 4.52$\pm$0.15 & 10.65 & $? $ \\
253 & J0957-0012 & C & 09\,57\,49.53 & -00\,12\,52.6 & 0.033 & 36.0$\pm$2.2 & 0.86$\pm$0.15 & 10.0 & $... $ \\
267 & J1008+0936 & C & 10\,08\,16.22 & +09\,36\,16.2 & 0.027 & 35.7$\pm$1.7 & ... & 9.97 & $... $ \\
268 & J1008+1916 & I & 10\,08\,28.72 & +19\,16\,19.9 & 0.182 & 14.3$\pm$1.0 & 2.31$\pm$0.14 & 10.96 & $... $ \\
270 & J1008+5123 & I & 10\,08\,47.68 & +51\,23\,52.8 & 0.156 & 14.1$\pm$1.1 & ... & 10.60 & $$\checkmark$ $ \\
293 & J1018+1536 & I & 10\,18\,23.97 & +15\,36\,30.9 & 0.111 & 36.0$\pm$1.5 & 2.99$\pm$0.15 & 10.78 & $? $ \\
305 & J1026+4340 & C & 10\,26\,53.35 & +43\,40\,08.4 & 0.105 & 41.9$\pm$1.5 & 1.30$\pm$0.15 & 10.25 & $$\checkmark$ $ \\
308 & J1028+5736 & I & 10\,28\,25.80 & +57\,36\,09.0 & 0.073 & 10.8$\pm$1.1 & 1.22$\pm$0.16 & 10.19 & $... $ \\
322 & J1031+0540 & I & 10\,31\,34.84 & +05\,40\,57.3 & 0.163 & 10.6$\pm$1.4 & ... & 10.72 & $$\checkmark$ $ \\
349 & J1046+2804 & I & 10\,46\,36.52 & +28\,04\,34.6 & 0.128 & 10.8$\pm$1.3 & 2.44$\pm$0.13 & 10.42 & $$\checkmark$ $ \\
365 & J1057+0554 & I & 10\,57\,51.07 & +05\,54\,46.8 & 0.054 & 27.7$\pm$1.5 & 1.82$\pm$0.15 & 10.06 & $... $ \\
437 & J1126+1913 & C & 11\,26\,19.44 & +19\,13\,29.2 & 0.103 & 48.8$\pm$1.6 & 3.57$\pm$0.15 & 10.48 & $$\checkmark$ $ \\
439 & J1127+1256 & C & 11\,27\,03.64 & +12\,56\,55.3 & 0.152 & 11.6$\pm$0.60 & 2.24$\pm$0.14 & 10.87 & $... $ \\
462 & J1136+2453 & C & 11\,36\,55.20 & +24\,53\,25.4 & 0.033 & 133.4$\pm$4.98 & 2.55$\pm$0.14 & 10.12 & $... $ \\
470 & J1139+4631 & C & 11\,39\,39.33 & +46\,31\,32.1 & 0.174 & 26.2$\pm$1.1 & 4.95$\pm$0.14 & 11.05 & $$\checkmark$ $ \\
498 & J1153+0930 & C & 11\,53\,41.32 & +09\,30\,25.5 & 0.139 & 48.5$\pm$2.1 & 1.56$\pm$0.21 & 10.75 & $... $ \\
533 & J1211+2936 & I & 12\,11\,38.23 & +29\,36\,16.5 & 0.107 & 14.1$\pm$1.1 & 1.32$\pm$0.13 & 10.59 & $... $ \\
547 & J1216+1904 & I & 12\,16\,22.27 & +19\,04\,42.2 & 0.075 & 11.7$\pm$1.1 & 5.76$\pm$0.14 & 10.90 & $$\checkmark$ $ \\
578 & J1229+3224 & C & 12\,29\,06.93 & +32\,24\,17.6 & 0.173 & 29.4$\pm$1.2 & 2.97$\pm$0.13 & 11.09 & $? $ \\
619 & J1248+5514 & I & 12\,48\,22.17 & +55\,14\,52.0 & 0.083 & 13.1$\pm$0.88 & ... & 10.75 & $... $ \\
658 & J1313+0207 & I & 13\,13\,52.39 & +02\,07\,57.3 & 0.030 & 12.9$\pm$1.2 & ... & 10.42 & $... $ \\
662 & J1314+2106 & C & 13\,14\,47.61 & +21\,06\,26.2 & 0.046 & 21.5$\pm$1.8 & 1.97$\pm$0.13 & 10.47 & $... $ \\
663 & J1315+2437 & C & 13\,15\,03.50 & +24\,37\,07.6 & 0.013 & 725.7$\pm$15.6 & 32.33$\pm$0.13 & 10.11 & $... $ \\
689 & J1326+1922 & C & 13\,26\,48.12 & +19\,22\,45.8 & 0.174 & 23.1$\pm$1.3 & 2.09$\pm$0.14 & 10.89 & $$\checkmark$ $ \\
704 & J1336+3008 & C & 13\,36\,04.12 & +30\,08\,27.9 & 0.026 & 39.4$\pm$2.3 & ... & 9.75 & $... $ \\
711 & J1339+4422 & C & 13\,39\,53.18 & +44\,22\,36.8 & 0.063 & 29.9$\pm$1.1 & ... & 10.47 & $$\checkmark$ $ \\
745 & J1356+2816 & C & 13\,56\,43.46 & +28\,16\,21.3 & 0.133 & 5.1$\pm$0.2 & 0.99$\pm$0.14 & 10.79 & $$\checkmark$ $ \\
767 & J1409+1016 & I & 14\,09\,52.53 & +10\,16\,46.9 & 0.096 & 4.5$\pm$0.2 & 2.32$\pm$0.15 & 10.87 & $$\checkmark$ $ \\
859 & J1505+5847 & I & 15\,05\,41.59 & +58\,47\,18.9 & 0.145 & 11.1$\pm$0.69 & 0.85$\pm$0.14 & 10.80 & $... $ \\
862 & J1506+0806 & I & 15\,06\,19.17 & +08\,06\,42.4 & 0.040 & 19.4$\pm$0.89 & ... & 10.24 & $... $ \\
908 & J1529+0601 & I & 15\,29\,08.37 & +06\,01\,19.5 & 0.106 & 18.3$\pm$0.97 & ... & 10.69 & $... $ \\
909 & J1529+0913 & I & 15\,29\,26.64 & +09\,13\,25.3 & 0.127 & 11.1$\pm$0.78 & ... & 10.49 & $$\checkmark$ $ \\
955 & J1555+2955 & I & 15\,55\,24.93 & +29\,55\,50.8 & 0.070 & 20.6$\pm$0.92 & 2.96$\pm$0.14 & 10.32 & $... $ \\
980 & J1611+0840 & I & 16\,11\,19.39 & +08\,40\,32.5 & 0.166 & 11.7$\pm$1.1 & ... & 10.59 & $... $ \\
1014 & J1645+3048 & I & 16\,45\,03.79 & +30\,48\,02.1 & 0.059 & 35.2$\pm$1.4 & ... & 9.98 & $... $ \\
1057 & J2245+1232 & I & 22\,45\,32.76 & +12\,32\,36.2 & 0.093 & 16.4$\pm$1.2 & ... & 10.70 & $$\checkmark$ $ \\
1062 & J2326-0114 & I & 23\,26\,37.22 & -01\,14\,36.2 & 0.197 & 38.4$\pm$1.7 & ... & 10.83 & $$\checkmark$ $ \\

\hline \hline
\end{tabular} \\
\label{tab:spogsCOsample}
\begin{minipage}{0.8\textwidth}
\raggedright{\footnotesize Column (1): SPOG \#. Column (2): SDSS name. Column (3): Telescope (I: IRAM\,30m, C: CARMA). Columns (4-5): SDSS RA/declination. Column (6): SDSS redshift. Column (7): 22$\mu$m flux from WISE detections. Column (8): 1.4\,GHz integrated flux density from FIRST. Column (9): Log of the stellar mass. Column (10): Morphological classification of each CO-SPOG as clearly disrupted ($\checkmark$), possibly disrupted (?), or not.  \\
}
\end{minipage}
\end{table*}

\addtocounter{footnote}{-1}
The Shocked POststarburst Galaxy Survey (SPOGS; \citealt{a16_sample})\footnote{\href{http://www.spogs.org}{http://www.spogs.org}} was created to search for rapidly transitioning galaxies that would be missed by poststarburst searches, aiming to identify galaxies that are quenching (rather than simply fading; \citealt{schawinski+14}). The SPOG sample was drawn from the Oh-Sarzi-Schawinski-Yi sample (OSSY; \citealt{ossy}), selecting only galaxies with bright emission in all diagnostic lines \citep{bpt,veilleux+87}, to create the parent Emission Line Galaxy sample (ELG; \citealt{a16_sample}).  The SPOGS criteria was applied to the ELG sample to include strong Balmer absorption (\ewhd)\footnote{It is possible that the SPOGS selection has missed shocked galaxies without ongoing star formation based on the requirement for such deep absorption, given the possibility of Balmer emission filling the stellar absorption features.} and ionized gas emission that is {\em inconsistent} with pure star formation. While SPOGS is by no means complete, these criteria have resulted in selecting 1,067 candidate objects (deemed SPOGs). Further details of the SPOG Survey are available in \citet{a16_sample}.

\begin{figure*}[t!]
\centering
\includegraphics[width=0.99\textwidth]{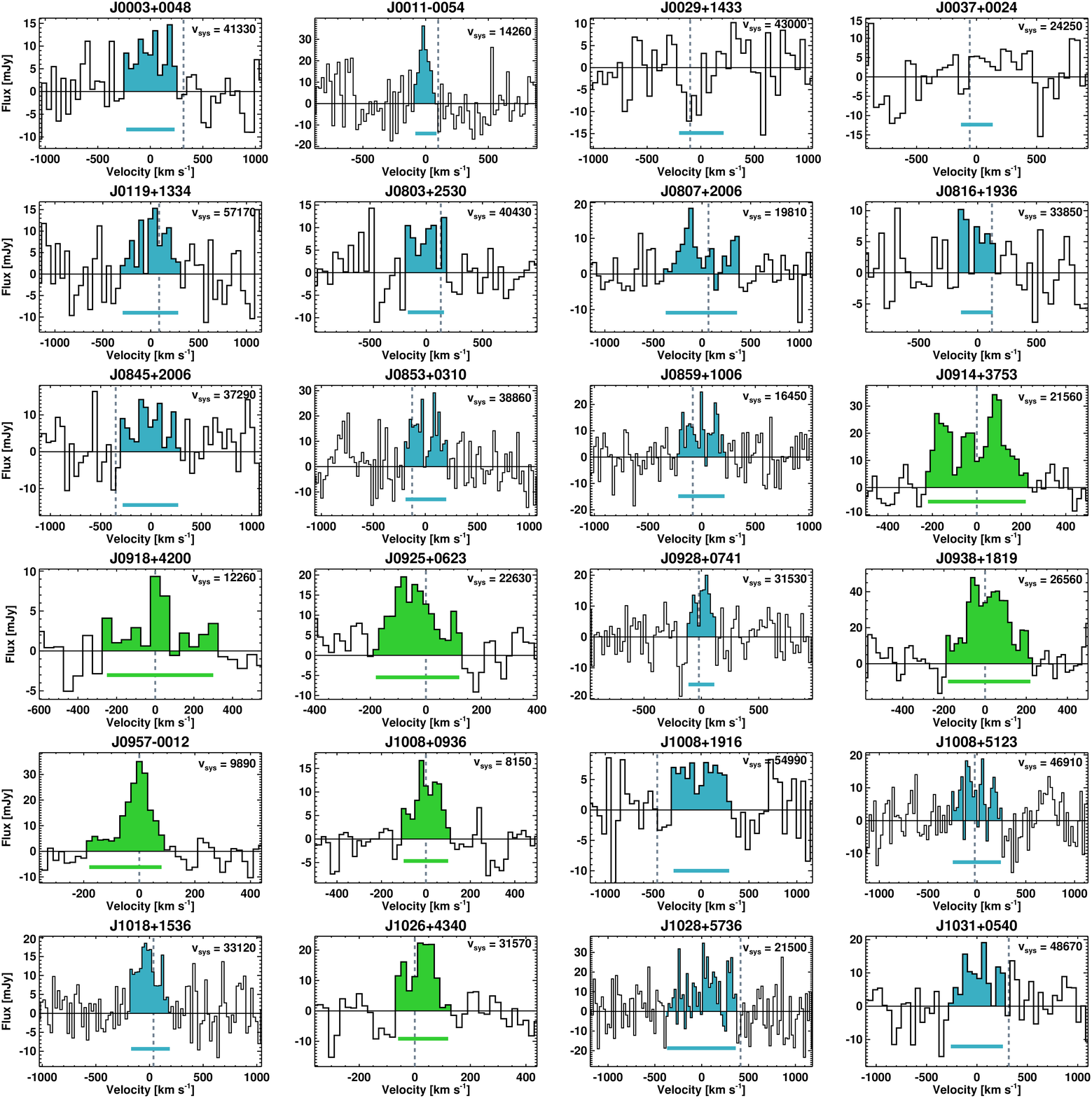}
\caption{The CO(1--0) spectra of the 52 observed SPOGs.  The IRAM~30m spectra are shaded in turquoise and CARMA in green.  A bar below the spectrum shows the velocity range used to sum over the CO(1--0) line. IRAM~30m objects with strong detections have 21\,km~s$^{-1}$ bins and those with more tentative detections with 42\,km~s$^{-1}$.  The optically-determined recession velocities are demarcated by a dotted gray line. The 5 non-detections (all from the IRAM~30m) were not shaded, but have velocity width bars to denote what velocity range is taken to calculate the upper limit. CARMA detections have velocity widths (which vary from source to source) noted in Table\,\ref{tab:carma}. The systemic velocities associated with each line are listed in the top corner of each panel.}
\label{fig:specs}
\end{figure*}
\renewcommand{\thefigure}{\arabic{figure} (Cont)}
\addtocounter{figure}{-1}
\begin{figure*}[t!]
\centering
\includegraphics[width=0.99\textwidth]{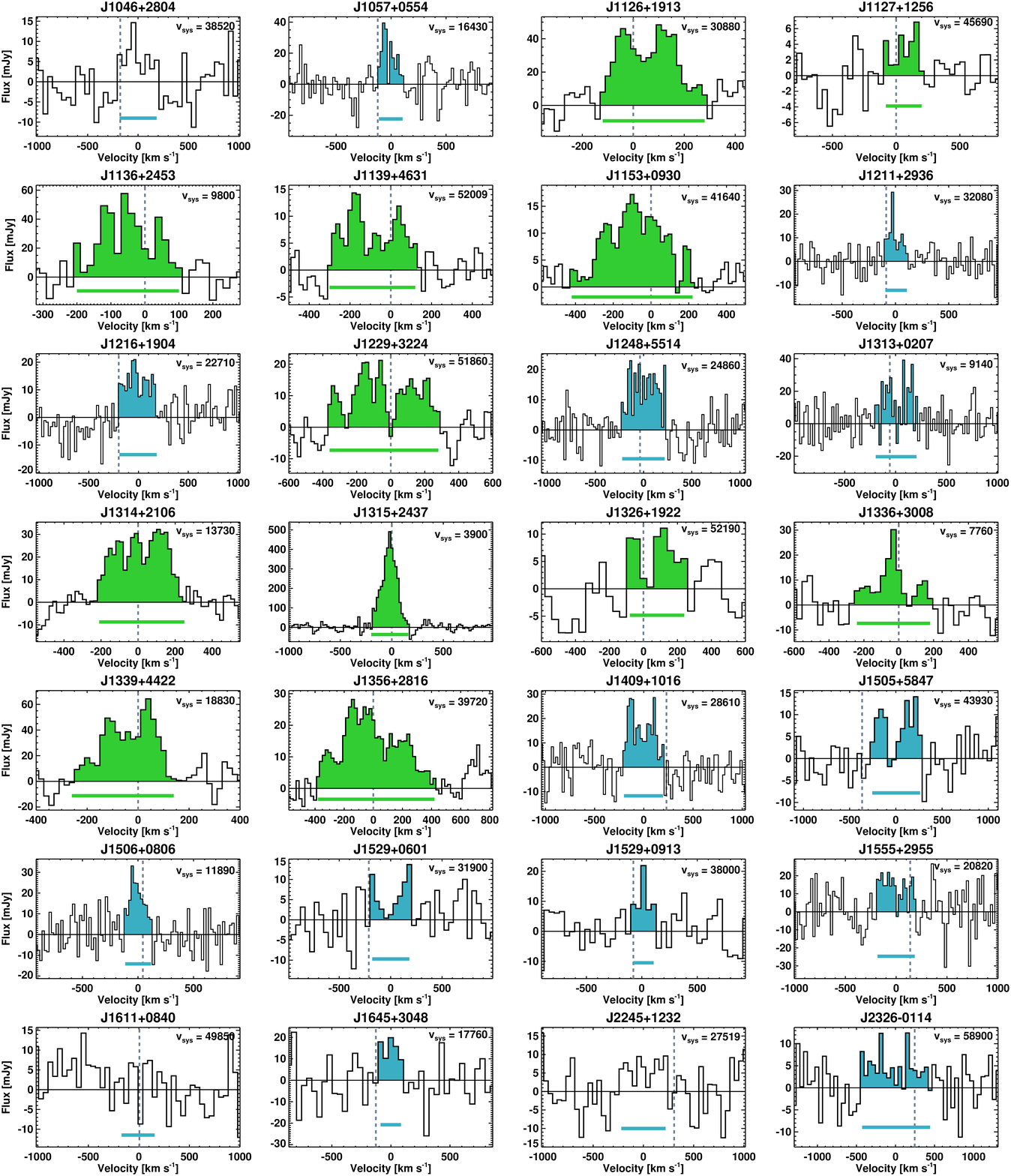}
\caption{Continuation of the IRAM~30m and CARMA CO(1--0) spectra}
\end{figure*}
\renewcommand{\thefigure}{\arabic{figure}}

We present new Combined Array from Research in Millimeter Astronomy (CARMA) and Institut de RAdioastronomie Millim\'etrique (IRAM) 30m CO(1--0) observations of 52 SPOGs.  In \S\ref{sec:sample}, we describe the selection used to draw the CO(1--0) sample. In \S\ref{sec:obs}, we describe the observations from IRAM and CARMA, including reduction and analysis methods.  In \S\ref{sec:results}, we present the molecular properties of the sample.  In \S\ref{sec:discussion}, we discuss these results in the context of transitioning galaxies.  In \S\ref{sec:summary}, we summarize our results.  The cosmological parameters $H_0$\,=\,70\,km~s$^{-1}$, $\Omega_{\rm m}$\,=\,0.3 and $\Omega_\Lambda$\,=\,0.7 \citep{wmap} are used throughout.


\section{The CO(1--0) Sample}
\label{sec:sample}
The objects chosen for CO(1--0) observations were selected based on the SPOG subsample cross-correlated with the {\em WISE} All-sky Survey \citep{wise}, detailed in \citet{a14_irtz}, to have detectable (signal-to-noise ratio $>$\,3) 22$\mu$m fluxes.  22$\mu$m emission is usually associated with star formation \citep{calzetti+07}, but it is also strong in AGNs \citep{ward+87,sanders+89,elvis+94}.  22$\mu$m emission can also arise in quiescent galaxies, from dust that is heated by the aging stellar population \citep{draine+07,crocker+11,crocker+13,dale+12}. Because our SPOG selection criteria removes galaxies whose ionized gas line ratios are dominated by star formation, the 22$\mu$m emission in these sources is less likely to be associated primarily with star formation. It is possible that some of our objects are in fact ``skin effect contaminants,'' in which the bulk of star formation is obscured from the optical view (in a compact core), and in which the overlying material is heated primarily by older stellar populations \citep{wild+11}. This cannot be ruled completely out until other star formation indicators have been measured. Despite this, it is unlikely that this is a dominant effect, as in the vast majority of cases, the optical emission and ionized gas line ratios of dusty, buried starbursts manifest as star-forming \citep{casey+14,rich+15}.  A caveat to this is that this scenario cannot remove star-forming objects completely and thus does not remove objects in which the dominant star formation is taking place outside of the 3$''$ SDSS fiber (although this is most problematic at low redshifts, where the fiber only traces the nucleus; \citealt{rich+14,a16_sample}).

Of the 1,067 SPOGs, 491 (46$\pm$2\%) are detected with a signal-to-noise ratio of at least 3 in the {\em WISE} 22$\mu$m band. Radio identifications of the {\em WISE} 22$\mu$m-detected sample were made via a 1.5$''$ radial match \citep{ivezic+02} with the Faint Images of the Radio Sky at Twenty Centimeters (FIRST) Survey \citep{first}, described in \citet{a16_sample}. There are 83/1,067 concurrent {\em WISE} 22$\mu$m and FIRST-detected objects (8$\pm$1\%), which accounts for over 50\% of the (160/1,067) total radio matches amongst the entire SPOG sample.

The IRAM~30m sample was selected from the 491 {\em WISE} 22$\mu$m-detected SPOGs. The {\em WISE} 22$\mu$m-detected sample was divided into bins based on radio detections and radio non-detections, as well as shock models within the \oi/\ha\ vs. \oiii/\hb\ line diagnostic diagram (\citealt{allen+08}; and Fig.~\ref{fig:wise_bpts}c), selecting one radio detection and one non-detection per bin.  Of the 40 IRAM\,30m proposed objects, 35 were observed and 30 were detected.  The CARMA-observed SPOG sample focused on a {\em WISE} 22$\mu$m flux-limited sample within the RA range of $9^h$--$15^h$.  There were 51 SPOGs with F$_{\rm 22\mu m}$\,$>$\,14\,mJy within this RA range that were queued for observation, and 19 were successfully observed. Table~\ref{tab:spogsCOsample} presents the general properties of the objects observed with the IRAM~30m and CARMA.

Our requirement for a {\em WISE} 22$\mu$m detection used to construct our follow-up sample of CO observations may bias this subset of SPOGs to favor those harboring AGNs, an aging stellar population, or dusty compact starbursts.  This is supported by the high prevalence of FIRST radio detections, which trace emission from AGNs (although can also trace star formation; \citealt{condon92}). Figure\,\ref{fig:wise_bpts} displays SPOGs (green circles) on the emission line ratio diagnostic diagrams \citep{bpt,veilleux+87}, with 22$\mu$m-detected sources (sky blue stars) identified, as well as concurrently 22$\mu$m+1.4\,GHz-detected objects (dark blue stars). Figure~\ref{fig:wise_bpts} shows that the emission line ratios of our CO-observed SPOG sample (hereafter, CO-SPOGs) do not exclusively lie in the Seyfert ionized gas ratio space or star-forming space \citep{kewley+06}. A large set of objects in this sample are consistent with the low ionization nuclear emission line region (LINER) portion of the diagram. This strongly suggests that the 22$\mu$m selection has selected AGNs in the SPOG sample. 

To improve accuracy, we calculate the total stellar masses of our CO-SPOGs by fitting the complete ({\em FUV, NUV, u, g, r, i, z, J, H, K}$_s$) photometry using {\sc magphys} \citep{magphys}, which takes into account extinction and k-corrections. SED fitting of the full SPOG sample will be presented in T. Bitsakis et al. (2016), in preparation.

The selection for the CO-SPOG sample also spans the entire {\em u--r} color-range of SPOGs (Fig.~\ref{fig:co_cmd}), but appear to trace more massive objects than the underlying SPOG population.

Between the IRAM~30m and CARMA samples, a total of 52 objects were observed (2 objects overlapped between the CARMA and IRAM~30m samples, with CARMA detections favored due to better signal-to-noise).  Figure\,\ref{fig:thumbs} shows the SDSS {\em g\,r\,i} 3-color thumbnails of all objects, and details of the observations are presented below.

\section{Observations and Analysis}
\label{sec:obs}
\subsection{The IRAM~30m}
We observed the CO(1--0)  line at the central position of our sample galaxies on September 9-14, October 15-18 and November 6-9, 2014 with the IRAM~30m telescope on Pico Veleta. We used the dual polarization Eight MIxer Receiver ({\sc emir}; \citealt{emir}) in combination with the autocorrelator Fourier Transform Spectrometers (FTS) at a frequency resolution of 0.195\,MHz at CO(1--0) (providing a velocity resolution of 0.57\,km~s$^{-1}$).  The observations were done in wobbler switching mode with a wobbler throw of 240\,\arcsec\ in the azimuthal direction.

\begin{table*}[t!]
\centering
\caption{CARMA Observational Parameters} \vspace{-1mm}
\begin{tabular}{l c c r r c c c}
\hline \hline
\# & SPOG & Total & Gain~~~~ & $\theta_{\rm maj}$$\times$$\theta_{\rm min}$ & $\Delta v$ & rms & Area\\
& Name & Hrs & Calibrator & ($''$)~~~~~ & (km~s$^{-1}$) & (mJy\,bm$^{-1}$) & ($\Box''$)\\
(1) & (2) & (3) & (4)~~~~~ & (5)~~~~~ & (6) & (7) & (8)\\
\hline
186 & J0914+3753 & 2.98 & 0927+390 & $10.1\times6.3$ & 20 & 1.4 & 222\\
191 & J0918+4200 & 5.44 & 0927+390 & $5.5\times4.4$ & 50 & 1.0 & 19\\
200 & J0925+0623 & 3.16 & 0825+031 & $8.4\times6.9$ & 20 & 1.4 & 142\\
224 & J0938+1819 & 2.11 & 0956+252 & $7.7\times7.0$ & 20 & 1.9 & 324\\
253 & J0957-0012 & 3.35 & 1058+015 & $8.7\times7.4$ & 20 & 1.3 & 125\\
267 & J1008+0936 & 2.44 & 1058+015 & $7.1\times6.6$ & 20 & 1.2 & 71\\
305 & J1026+4340 & 2.17 & 0927+390 & $9.2\times6.5$ & 20 & 1.5 & 242\\
437 & J1126+1913 & 1.85 & 1159+292 & $9.9\times6.4$ & 20 & 1.7 & 290\\
439 & J1127+1256 & 3.37 & 1058+015 & $8.4\times7.3$ & 40 & 2.7 & 67\\
462 & J1136+2453 & 0.82 & 1224+213 & $4.1\times3.5$ & 20 & 4.7 & 36\\
470 & J1139+4631 & 3.63 & 1153+495 & $8.4\times6.9$ & 20 & 0.8 & 180\\
498 & J1153+0930 & 6.69 & 3C273 & $9.3\times6.8$ & 20 & 0.6 & 133\\
578 & J1229+3224 & 5.82 & 1159+292 & $9.9\times6.7$ & 20 & 0.9 & 245\\
662 & J1314+2106 & 4.02 & 1310+323 & $7.9\times6.1$ & 20 & 1.4 & 232\\
663 & J1315+2437 & 4.97 & 1310+323 & $2.6\times2.1$ & 20 & 15.1 & 134\\
689 & J1326+1922 & 2.84 & 1310+323 & $8.2\times7.5$ & 40 & 0.9 & 201\\
704 & J1336+3008 & 4.99 & 1310+323 & $7.2\times5.9$ & 30 & 1.3 & 172\\
711 & J1339+4422 & 2.41 & 1419+543 & $8.5\times6.1$ & 20 & 2.2 & 343\\
745 & J1356+2816 & 5.03 & 1310+323 & $8.5\times6.5$ & 20 & 0.8 &       48\\

\hline \hline
\end{tabular} \\
\label{tab:carma}
\begin{minipage}{0.58\textwidth}
\raggedright{\footnotesize Column (1): SPOG number. Column (2): SDSS name. Column (3): Total time on-source. Column (4): Gain calibrator used. Column (5): CARMA beam full-width at half maximum. Column (6): Channel velocity width. Column (7): Intensity RMS per beam of CARMA images. Column (8): Moment0 aperture area.
}
\end{minipage}
\end{table*}

The broad bandwidth of the receiver (16\,GHz) and backend allowed us to group the observations of galaxies with similar redshifts. The central sky frequencies, taking into account the redshift of the objects, ranged between 96 and 111\,GHz.  Each object was observed until it was detected with a S/N  ratio of at least 5 or until a RMS of 1.5\,mK (T$_{\rm A}^*$) was achieved for a velocity resolution of 20\,km~s$^{-1}$. The integration times per object ranged between half an hour and 2 hours, with a mean value of 80 minutes.  Pointing was monitored on nearby quasars  every 60-90 minutes.  During the observation period, the weather conditions were generally good, with a pointing accuracy better than 4\arcsec. The mean system temperature for the observations was 116\,K on the $T_{\rm A}^*$ scale.  At 115 GHz, the IRAM forward efficiency, $F_{\rm eff}$, was 0.95; the beam efficiency, $B_{\rm eff}$, was 0.79; and the half-power beam size ranges between 22.3$^{\prime\prime}$ (for 110\,GHz)  and 25.6$^{\prime\prime}$ (for 97\,GHz).  All CO spectra and luminosities are presented on the Jansky scale, converted from the main beam temperature scale ($T_{\rm mb}$) which is defined as $T_{\rm mb} = (F_{\rm eff}/B_{\rm eff})\times T_{\rm A}^*$, using a conversion factor of 5 Jy/K.

\begin{table*}[t!]
\centering
\caption{SPOG CO(1--0) Values} \vspace{-1mm}
\begin{tabular}{l c r c r r r r r c}
\hline \hline
\# & SPOG & Vel. Range~~& RMS & F$_{\rm CO}$~~~~~~ & SNR$_{\rm CO}$ & L$_{\rm CO}$~~~~~~~  & log($M_{\rm H_2}$)~~ & $F_{\rm gas}$  \\
& Name & (km~s$^{-1}$)~~~ & (mJy) & (Jy~km~s$^{-1}$) & & (10$^4~L_\odot$)~~ & ($M_\odot$)~~~~ \\
(1) & (2) & (3)~~~~~~~~ & (4)~~ & (5)~~~~~~~ & (6) & (7)~~~~~~~~ & (8)~~~~~~ & (9)~~~\\
\hline
1 & J0003+0048 & 41070--41580 & 7.20 & $4.19\pm0.78$ & 5.4 & $18.99\pm3.53$ & $10.22\pm9.5$ & 0.24 \\
4 & J0011-0054 & 14150--14350 & 8.80 & $3.05\pm0.58$ & 5.3 & $1.58\pm0.30$ & $9.14\pm8.4$ & 0.06 \\
7 & J0029+1433 & 42800--43200 & 9.05 & $<7.80 $ & -- & $<37.58 $ & $<10.52 $ & 0.40 \\
13 & J0037+0024 & 24100--24400 & 7.45 & $<5.45 $ & -- & $<8.13 $ & $<9.85 $ & 0.32 \\
24 & J0119+1334 & 57030--57480 & 8.70 & $3.94\pm0.92$ & 4.3 & $34.41\pm7.99$ & $10.48\pm9.8$ & 0.25 \\
77 & J0803+2530 & 40250--40620 & 7.80 & $2.86\pm0.73$ & 3.9 & $12.28\pm3.11$ & $10.03\pm9.4$ & 0.11 \\
81 & J0807+2006 & 19590--20190 & 7.05 & $3.79\pm0.80$ & 4.7 & $3.79\pm0.80$ & $9.52\pm8.8$ & 0.12 \\
98 & J0816+1936 & 33670--33980 & 6.15 & $2.01\pm0.52$ & 3.9 & $5.98\pm1.54$ & $9.72\pm9.1$ & 0.17 \\
142 & J0845+2006 & 36960--37530 & 10.05 & $4.79\pm1.16$ & 4.1 & $16.95\pm4.09$ & $10.17\pm9.6$ & 0.27 \\
157 & J0853+0310 & 38660--39080 & 7.80 & $5.42\pm0.77$ & 7.1 & $21.15\pm2.99$ & $10.27\pm9.4$ & 0.20 \\
169 & J0859+1006 & 16240--16670 & 7.65 & $4.15\pm0.74$ & 5.6 & $2.80\pm0.50$ & $9.39\pm8.6$ & 0.07 \\
186 & J0914+3753 & 21340--21780 & 1.40 & $7.48\pm0.13$ & 57.1 & $8.83\pm0.15$ & $9.89\pm8.1$ & 0.28 \\
191 & J0918+4200 & 12010--12560 & 1.04 & $1.70\pm0.17$ & 9.9 & $0.64\pm0.06$ & $8.75\pm7.8$ & 0.03 \\
200 & J0925+0623 & 22450--22750 & 1.42 & $3.30\pm0.11$ & 29.9 & $4.30\pm0.14$ & $9.58\pm8.1$ & 0.10 \\
209 & J0928+0741 & 31410--31680 & 6.20 & $2.47\pm0.49$ & 5.1 & $6.32\pm1.24$ & $9.74\pm9.0$ & 0.30 \\
224 & J0938+1819 & 26380--26780 & 1.88 & $10.40\pm0.17$ & 61.9 & $18.78\pm0.30$ & $10.22\pm8.4$ & 0.27 \\
253 & J0957-0012 & 9710--9970 & 1.30 & $3.75\pm0.09$ & 39.9 & $0.92\pm0.02$ & $8.90\pm7.3$ & 0.07 \\
267 & J1008+0936 & 8050--8250 & 1.17 & $1.88\pm0.07$ & 25.5 & $0.31\pm0.01$ & $8.43\pm7.0$ & 0.03 \\
268 & J1008+1916 & 54750--55270 & 6.60 & $3.21\pm0.74$ & 4.3 & $25.31\pm5.84$ & $10.35\pm9.7$ & 0.20 \\
270 & J1008+5123 & 46650--47170 & 6.60 & $3.84\pm0.73$ & 5.3 & $22.21\pm4.22$ & $10.29\pm9.6$ & 0.33 \\
293 & J1018+1536 & 32920--33320 & 5.05 & $4.12\pm0.48$ & 8.6 & $11.70\pm1.36$ & $10.01\pm9.1$ & 0.15 \\
305 & J1026+4340 & 31509--31690 & 1.47 & $2.51\pm0.09$ & 28.5 & $6.44\pm0.23$ & $9.75\pm8.3$ & 0.24 \\
308 & J1028+5736 & 21120--21880 & 10.95 & $7.32\pm1.40$ & 5.2 & $8.93\pm1.71$ & $9.89\pm9.2$ & 0.34 \\
322 & J1031+0540 & 48490--48980 & 9.55 & $4.66\pm1.02$ & 4.5 & $29.46\pm6.49$ & $10.41\pm9.8$ & 0.33 \\
349 & J1046+2804 & 38300--38740 & 8.45 & $<2.55 $ & -- & $<9.75 $ & $<9.93 $ & 0.24 \\
365 & J1057+0554 & 16299--16560 & 11.05 & $4.04\pm0.83$ & 4.9 & $2.70\pm0.56$ & $9.37\pm8.7$ & 0.17 \\
437 & J1126+1913 & 30760--31160 & 1.67 & $11.69\pm0.15$ & 78.1 & $28.69\pm0.37$ & $10.40\pm8.5$ & 0.45 \\
439 & J1127+1256 & 45610--45890 & 2.73 & $1.06\pm0.29$ & 3.7 & $5.81\pm1.58$ & $9.71\pm9.1$ & 0.06 \\
462 & J1136+2453 & 9600--9900 & 4.71 & $8.40\pm0.37$ & 23.0 & $2.01\pm0.09$ & $9.25\pm7.9$ & 0.12 \\
470 & J1139+4631 & 51710--52130 & 0.77 & $3.09\pm0.07$ & 43.9 & $22.10\pm0.50$ & $10.29\pm8.6$ & 0.15 \\
498 & J1153+0930 & 41220--41860 & 0.60 & $5.43\pm0.07$ & 80.3 & $24.61\pm0.31$ & $10.33\pm8.4$ & 0.28 \\
533 & J1211+2936 & 31970--32200 & 5.50 & $2.32\pm0.40$ & 5.9 & $6.12\pm1.04$ & $9.73\pm9.0$ & 0.12 \\
547 & J1216+1904 & 22510--22900 & 6.30 & $4.93\pm0.59$ & 8.4 & $6.35\pm0.75$ & $9.74\pm8.8$ & 0.07 \\
578 & J1229+3224 & 51500--52140 & 0.88 & $6.94\pm0.10$ & 69.3 & $49.38\pm0.71$ & $10.64\pm8.8$ & 0.26 \\
619 & J1248+5514 & 24610--25090 & 5.20 & $6.25\pm0.54$ & 11.7 & $9.82\pm0.84$ & $9.93\pm8.9$ & 0.13 \\
658 & J1313+0207 & 8920--9360 & 10.45 & $5.50\pm1.01$ & 5.5 & $1.13\pm0.21$ & $8.99\pm8.3$ & 0.04 \\
662 & J1314+2106 & 13520--13980 & 1.37 & $9.97\pm0.13$ & 75.9 & $4.71\pm0.06$ & $9.62\pm7.7$ & 0.12 \\
663 & J1315+2437 & 3700--4059 & 15.14 & $76.82\pm1.28$ & 59.8 & $2.88\pm0.05$ & $9.40\pm7.6$ & 0.16 \\
689 & J1326+1922 & 52110--52430 & 0.88 & $2.34\pm0.10$ & 23.5 & $16.89\pm0.72$ & $10.17\pm8.8$ & 0.16 \\
704 & J1336+3008 & 7530--7950 & 1.35 & $3.97\pm0.15$ & 26.3 & $0.59\pm0.02$ & $8.72\pm7.3$ & 0.08 \\
711 & J1339+4422 & 18570--18970 & 2.23 & $12.16\pm0.20$ & 60.9 & $10.90\pm0.18$ & $9.98\pm8.2$ & 0.24 \\
745 & J1356+2816 & 39340--40140 & 0.77 & $11.37\pm0.10$ & 117.4 & $46.77\pm0.40$ & $10.61\pm8.5$ & 0.40 \\
767 & J1409+1016 & 28390--28809 & 6.85 & $6.14\pm0.66$ & 9.3 & $13.11\pm1.41$ & $10.06\pm9.1$ & 0.13 \\
859 & J1505+5847 & 43630--44220 & 8.15 & $4.04\pm0.96$ & 4.2 & $20.06\pm4.77$ & $10.24\pm9.6$ & 0.22 \\
862 & J1506+0806 & 11770--12010 & 8.75 & $4.29\pm0.63$ & 6.8 & $1.53\pm0.22$ & $9.13\pm8.3$ & 0.07 \\
908 & J1529+0601 & 31669--32130 & 7.40 & $2.79\pm0.75$ & 3.7 & $7.22\pm1.94$ & $9.80\pm9.2$ & 0.11 \\
909 & J1529+0913 & 37900--38140 & 8.45 & $2.52\pm0.63$ & 4.0 & $9.42\pm2.34$ & $9.92\pm9.3$ & 0.21 \\
955 & J1555+2955 & 20610--21010 & 11.65 & $5.10\pm1.08$ & 4.7 & $5.68\pm1.20$ & $9.70\pm9.0$ & 0.19 \\
980 & J1611+0840 & 49700--50000 & 11.60 & $<8.75 $ & -- & $<57.40 $ & $<10.70 $ & 0.56 \\
1014 & J1645+3048 & 17620--17850 & 14.45 & $3.35\pm1.02$ & 3.3 & $2.63\pm0.80$ & $9.36\pm8.8$ & 0.19 \\
1057 & J2245+1232 & 27400--27750 & 9.35 & $<2.46 $ & -- & $<4.88 $ & $<9.63 $ & 0.08 \\
1062 & J2326-0114 & 58470--59340 & 8.70 & $4.51\pm1.27$ & 3.6 & $42.07\pm11.8$ & $10.57\pm10.0$ & 0.35 \\

\hline \hline
\end{tabular} \\
\label{tab:spogsCO}
\begin{minipage}{0.75\textwidth}
\raggedright {\footnotesize  Column (1): SPOG \#. Column (2): SDSS name. Column (3): Velocity range (optically-defined, local standard of rest). Column (4): Spectral RMS. Column (5): CO(1--0) line flux. Column (6): The signal-to-noise ratio of the CO detections. Column (7): CO(1--0) luminosity. Column (8): Observed mass of H$_2$ (using the conversion factor of \citealt{bolatto+13}). Column (9): Molecular gas fraction, $F_{\rm gas} = \frac{M_{\rm H_2}}{M_{\rm H_2}+M_{\rm star}}$ \\
}
\end{minipage}
\end{table*}

For the data reduction, we first discarded poor scans and then subtracted a constant or linear baseline. A large number of scans were affected by platforming, i.e. the baseline level changed abruptly at one or two positions along the band. This effect could be reliably corrected because the baselines in between these (clearly visible) jumps were flat and allowed to determine the (order 0 or 1) baseline which had to be subtracted from  the different parts in order to move the baselines to a zero level along the entire band.  After this correction we summed the spectra of each source, smoothed them to resolutions between 20--40\,km~s$^{-1}$ in order to increase the S/N ratio per channel, and visually determined the zero-level widths (the boundaries beyond which the spectra drop to zero). CO-SPOGs with S/N ratios $>$3 are considered detected, and CO-SPOGs with S/N ratios $>$5 are considered strongly detected.  Figure~\ref{fig:specs} shows the integrated spectra of the 35 SPOGs observed with the IRAM~30m (shaded turquoise). The systemic velocity of the line is set to zero, and the total linewidth is represented by a turquoise line and is shaded in the spectrum. The optically-defined systemic velocity is shown as a black dotted line. The velocity integrated spectra were calculated by summing the individual shaded channels and multiplying by the velocity width of each channel (21 or 42\,km~s$^{-1}$). 

As an alternative approach, we fit the spectra with a Gaussian profile and integrated it over velocity (details of the Gaussian fitting can be found in \S\ref{app:gauss}). The velocity integrated intensity determined by the two methods are in good agreement, with a mean difference of 5\%, confirming that our velocity integration is reliable.  The line flux RMS was then calculated by multiplying the RMS per channel by the channel velocity width and the square root of the total number of channels determined to contain line emission (those that are shaded).

\subsection{CARMA}
The CARMA SPOG observations were taken between June and December 2014 using CARMA, an interferometer of 15 radio dishes (6$\times$10.4m and 9$\times$6.1m) located in the Eastern Sierras in California \citep{carma}\footnote{\href{http://www.mmarray.org}{http://www.mmarray.org}}. 19 SPOGs were observed (with a 100\% detection rate), either in D- or E-array, with baselines between 11--150\,m and 8--66\,m, respectively. Standard reduction and calibration techniques (as described in detail in \citealt{alatalo+13}) were used on all targets.

Our SPOGs were unresolved in all cases except J1315+2437 (IC\,860, which was also observed in CARMA C-array by \citealt{mcbride+14}).  The observing parameters associated with each of the CARMA SPOGs are listed in Table\,\ref{tab:carma}. A moment0 map for each CARMA SPOG (using the {\sc miriad} task {\tt moment}; \citealt{miriad}), in which a sigma clip was applied to velocity channels determined to have emission (see \citealt{alatalo+13} for details). We did not apply a standard sigma clip to the channel maps, instead iteratively determining the correct sigma value to maximize the detection of real emission to the exclusion of noise in each individual object. Figure\,\ref{fig:specs} shows the integrated spectra of the 19 SPOGs observed with CARMA (shaded green).  The systemic velocity of the line is set to zero, with the total linewidth represented by a green line, as well as in the shaded region of the spectrum. The integrated spectrum of each SPOG is determined by integrating the flux within an aperture determined by the moment0 map created for each CARMA SPOG.

The RMS per channel is calculated by (1) taking the standard deviation of all pixels within the cube that were outside the moment0 aperture, (2) applying an additional noise up-correction of 30\% to account for the oversampling of the maps (see: \citealt{a15_co13} for details), and (3) multiplying by the square root of the total number of beams represented in the moment0 aperture.

\begin{figure}[t]
\subfigure{\includegraphics[width=0.475\textwidth]{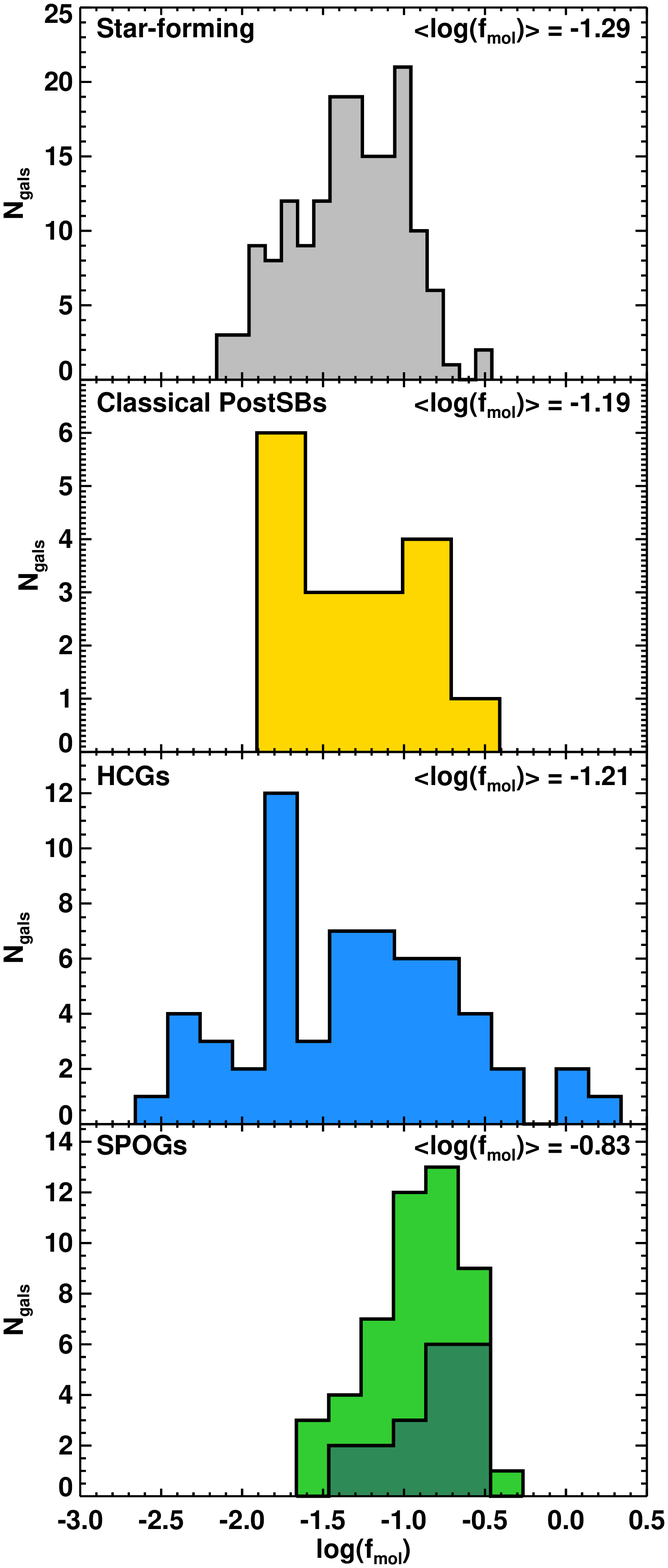}}
\vskip -3mm
\caption{The distribution of molecular gas to stellar mass ratio in the COLD GASS star-forming galaxies (top; \citealt{coldgass}), the classical poststarburst galaxies (middle; yellow; \citealt{french+15}), HCG galaxies (middle, blue; \citealt{leon+98,martinez-badenes+12,lisenfeld+14}, 2016 in prep) and CO-SPOGs (bottom; light or dark green), assuming the same $X_{\rm CO}$. The molecular-to-stellar mass fractions of CO-SPOGs are higher than those seen in both the star-forming galaxies and poststarbursts. All stellar masses were calculated as described in \S\ref{disc:fmol}. The distribution of morphologically disrupted SPOGs (dark green) is overlaid, and reflects the overall SPOG distribution faithfully.}
\label{fig:fmol}
\end{figure}

To calculate the integrated line flux for each galaxy, we summed the channels shaded in green in Fig.~\ref{fig:specs} and multiplied by the velocity width of each channel, listed in Table\,\ref{tab:carma}.  The line flux RMS was then calculated by multiplying the RMS per channel by the channel velocity width and the square root of the total number of channels determined to contain line emission. In the case of the two SPOGs observed by both the IRAM~30m and CARMA, the CO fluxes agreed to within the standard 20\% errors.

\section{Results}
\label{sec:results}
\subsection{The Morphologies of CO-SPOGs}
Figure~\ref{fig:thumbs} show the 3-color {\em g\,r\,i} images from SDSS of all 52 CO-SPOGs. The thumbnails show a large number of galaxies that include signs of interaction, though only a handful appear to be major mergers in the coalescence phase \citep{dopita+02}. A significant fraction of CO-SPOGs have tidal features prominent enough to be seen in SDSS images. The remaining sample of CO-SPOGs mainly consists of early-type spirals, lenticular galaxies with bars, and objects with peaked nuclei consistent with AGN. Our selection did not include morphology, so it is interesting that the number of tidally disrupted objects represented in CO-SPOGs is larger than that in the general SPOG sample.

In order to quantify whether the CO-SPOG galaxies are disrupted, six team members (Alatalo, Appleton, Cales, Lanz, Lisenfeld, Nyland) independently visually inspected the {\em g\,r\,i} thumbnails of these galaxies (Fig.~\ref{fig:thumbs}). Based on the (potential) presence of tidally features, they classified the galaxies as (possibly) disrupted or not. There was complete consensus of the classification of 47 galaxies, indicated by the presence or absence of a check mark in Table~\ref{tab:spogsCOsample}, as well as a (\textcolor{Dandelion}{$\mathbf{\infty}$}) symbol on Figure~\ref{fig:thumbs}. The other 5 galaxies had at least two classifications differing from the others and are therefore marked as ambiguous, represented as a question mark in Table~\ref{tab:spogsCOsample} and (\textcolor{Dandelion}{$\mathbf{\sim}$}) in Figure~\ref{fig:thumbs}. Based on these classifications, 19--24 (37--46$\pm$7\%\footnote{Errors are those from assuming a binomial distribution}) of our CO-SPOGs are classified as disrupted. A detailed analysis of the morphologies of SPOGs will be presented in a future paper.

\subsection{CO(1--0) Properties}
Figure~\ref{fig:specs} shows the CO(1--0) spectra of all CO-SPOGs, color coded to identify which facility was used to make the observation (turquoise for the IRAM~30m and green for CARMA).  A few double-horned spectral profiles, consistent with molecular gas rotating in a disk that extends out to the flat part of the rotation curve, are present but are the minority of detections (although this could in part be due to sensitivity).  In many of the strong molecular gas detections, multiple components and peaks are present in the molecular gas, consistent with the morphological disruption present in {\em g\,r\,i} images of the galaxies.

Table \ref{tab:carma} lists the derived properties for CARMA-observed CO-SPOGs, including the RMS noise in the channel maps, the spatial extent of the molecular gas (determined by summing the total number of unmasked pixels in the moment map; see \citealt{alatalo+13} for details of moment map construction), channel widths, and total on-source hours. Derived molecular gas properties are listed in Table~\ref{tab:spogsCO}. CO luminosities are derived using the equation in \citet{solomon+vandenbout05}:
\begin{equation}
L_{\rm CO} = 1.20\times10^{-1} \frac{S_{\rm CO} \Delta v~D_L^2}{1+v_{\rm sys}/c}~L_\odot,
\end{equation}

\noindent where $S_{\rm CO} \Delta v$ is the CO(1--0) flux (in Jy~km~s$^{-1}$), $D_{\rm L}$ is the luminosity distance (in Mpc), $v_{\rm sys}$ is the optically-defined systemic velocity (in km~s$^{-1}$), and $c$ is the speed of light (in km~s$^{-1}$). Converting the CO(1--0) luminosity into molecular gas masses requires the assumption of a \lcomhtwo\ conversion factor \citep{bolatto+13}, which is known to be dependent on the state of the molecular gas. Interacting galaxies and ultra-luminous infrared galaxies (ULIRGs) are known to have \lcomhtwo\ conversions that are lower than normal, star-forming galaxies by a factor of about 5 \citep{downes+solomon98}, though \citet{narayanan+11} have shown that there is a large scatter above and below the standard value, even within merging systems.  \citet{sandstrom+13} showed that the nuclei of normal star-forming galaxies can exhibit CO--to--H$_2$ conversion up to an order of magnitude below the Milky Way value. Determining the proper \lcomhtwo\ conversion is therefore essential to deriving the molecular mass of a galaxy.

\vspace{2mm}
\noindent We calculate $M_{\rm H_2}$ with the following equation: 

\begin{equation}
M_{\rm H_2} = 1.05\times10^4\frac{S_{\rm CO}\Delta v D_L^2}{1+v_{\rm sys}/c}~M_\odot
\end{equation}

\begin{figure*}[t]
\subfigure{\includegraphics[width=\textwidth]{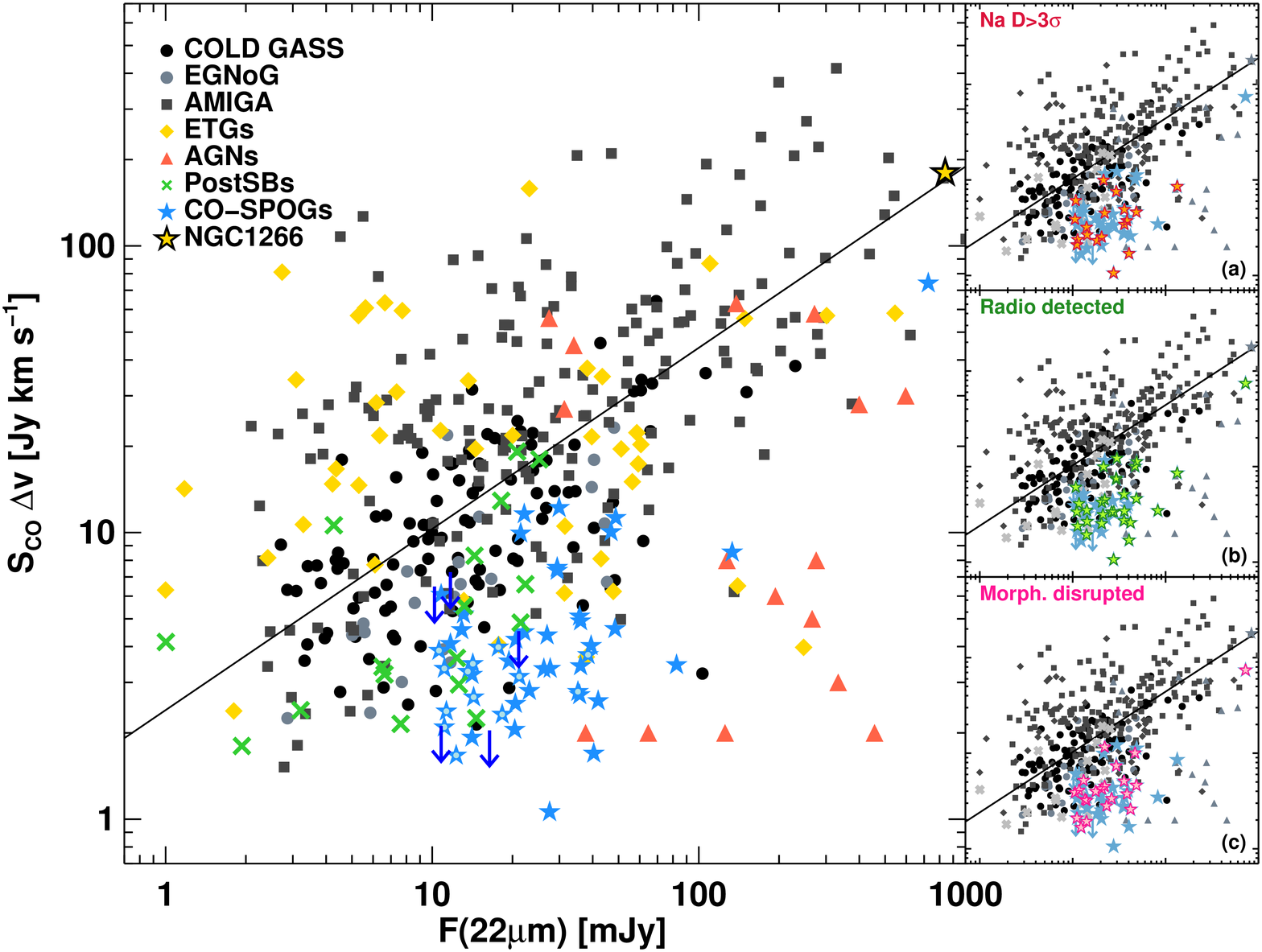}}
\caption{{\bf (Left):} The 22$\mu$m and CO(1--0) fluxes are compared for many samples, including star-forming galaxies such as COLD GASS (black dots; \citealt{coldgass}), EGNoG (gray dots; \citealt{bauermeister+13}) and AMIGA (dark gray squares; \citealt{lisenfeld+11}), early-type galaxies (yellow diamonds; ATLAS$^{\rm 3D}$: \citealt{young+11,alatalo+13,davis+14}), AGNs (red triangles; \citealt{evans+05}), poststarburst galaxies (green crosses; \citealt{french+15}), NGC\,1266 (outlined yellow star; \citealt{alatalo+11,a14_stelpop,a15_sfsupp}), and CO-SPOGs (blue stars). Light blue dots inside of the stars denote CO-SPOGs that have detections with S/N between 3--5. The 22$\mu$m and CO(1--0) fluxes of star-forming galaxies are well correlated (shown as a black line), with AGNs showing the strongest divergence from the relation.  CO-SPOGs sit below the relation as well, though not as extremely as the AGNs. {\bf (Right):} The 22$\mu$m vs. CO(1--0) fluxes are shown, emphasizing the SPOGs with particular properties (symbols for other samples from lefthand plot have been changed to grayscale). This includes those galaxies with 3$\sigma$ excess \nad\ absorption (a; top right, red), galaxies that were detected with FIRST (b; middle right, green), and galaxies that were classified as clearly or possibly morphologically disrupted (c; bottom right, pink). Although radio-detected CO-SPOGs account for most of the objects that fall the farthest from the $F_{22}$--$F_{\rm CO}$ relation, they do not universally exhibit quasar-like 22$\mu$m excess relative to their CO fluxes.}
\label{fig:fco22}
\end{figure*}

\noindent assuming $X_{\rm CO} = 2\times10^{20}$\,cm$^{-2}$~(K~km~s$^{-1}$)$^{-1}$, the mean conversion factor presented in \citet{bolatto+13},  the derived H$_2$ masses for our CO-SPOGs range between 10$^{8.4-10.6}\,M_\odot$.

\section{Discussion}
\label{sec:discussion}

\subsection{The molecular gas fraction of CO-SPOGs}
\label{disc:fmol}

Figure~\ref{fig:fmol} shows the molecular gas fraction distribution for different subsets of galaxies that contain molecular gas. Normal, star-forming galaxies from CO Legacy Database for the {\em GALEX} Arecibo SDSS Survey (COLD GASS; top; \citealt{coldgass}), have an average molecular gas fraction $\langle \log(f_{\rm mol,SF})\rangle = -1.29$. The classical poststarbursts (middle; \citealt{french+15}) have $\langle \log(f_{\rm mol,PSB})\rangle = -1.19$, and CO-SPOGs (bottom) have $\langle \log(f_{\rm mol,SPOG})\rangle = -0.83$. SPOGs tend to have higher gas fractions than both normal galaxies and poststarbursts. The molecular gas fractions in relation to the stellar masses range from 10$^{-1.56}$ to 10$^{-0.34}$, with those SPOGs at the high end of the molecular gas fraction range being comparable to interactions (in which log($M_{\rm H_2}/M_{\rm star}+M_{\rm H_2}$)\,$\sim$\,-0.3; \citealt{combes+13,kaneko+15}). We ran a Mann Whitney U-test\footnote{{\sc idl} routine: {\tt rs\_test}} (which is used in cases of small numbers), and were able to show that $f_{\rm mol}$ in SPOGs is distinct from both star-forming galaxies ({\em p}\,$\approx$\,0) as well as poststarbursts ({\em p}\,=\,0.0019), where {\em p} is the probability of a null hypothesis.

The fraction of our CO-SPOGs classified as having disrupted morphologies is large (37--46$\pm$7\%). This optical disruption suggests that the molecular gas should also be disrupted to some extent. \citet{downes+solomon98} showed that in many interacting systems, the molecular gas mass predicted from the CO(1--0) flux was larger than the dynamical mass of the system, usually by about a factor of 5, though this is quite uncertain \citep{yao+03,bolatto+13}. They conclude that this is due to the nature of the molecular gas in interacting systems being more diffuse and warm \citep{aalto+95,rangwala+11}, creating a more continuous coverage of CO-emitting gas rather than distributed in discrete giant molecular clouds (which would also span a smaller velocity range). If the molecular gas in SPOGs is indeed disrupted and therefore not confined to a disk as in the Milky Way, we could be overestimating the molecular gas mass by $\sim$5. If CO-SPOGs are found on the extreme end of the conversions found by \citet{sandstrom+13} for nuclear regions of star-forming galaxies, then the conversion could be off by as much as a factor of 7. If all CO-SPOGs require one of these reduced conversion factors, then the molecular-to-stellar mass ratio would shift significantly, to an average ratio of 10$^{-1.68}$, consistent with the COLD GASS sample \citep{coldgass}, and less than the poststarburst sample \citep{french+15}.

Figure~\ref{fig:fmol} also overlays the molecular gas fraction of a sample of Hickson Compact Group (HCG) galaxies, with stellar masses from U. Lisenfeld et al. 2016 in preparation, and H$_2$ masses calculated from CO observations \citep{leon+98,verdes-montenegro+98,martinez-badenes+12,lisenfeld+14}, using $X_{\rm CO}$\,=\,2$\times$\,10$^{20}$\,cm$^{-2}$~(K~km~s$^{-1}$)$^{-1}$. HCG galaxies to span the range of $f_{\rm mol}$ values, suggestive of their large range of properties and environments \citep{bitsakis+10,bitsakis+11,bitsakis+14,zucker+16}. Many HCGs contain warm H$_2$-luminous galaxies, known to host shocks \citep{cluver+13}, which may be a better comparison sample with CO-SPOGs than interactions. Given that the SPOG criteria aimed to select objects that host shocks, the warm H$_2$-bright galaxies serve as a reasonable comparison. \citet{a15_hcgco} used CARMA to study the molecular gas in turbulent HCG galaxies and found that the molecular gas-to-dust ratios within these systems were consistent with the Milky Way value, and predicted gas masses that were consistent with a normal CO-to-H$_2$ conversion. Despite the fact that both types of systems host shocks, the molecular-to-stellar mass ratio for warm H$_2$ bright HCG galaxies tends to be lower than for SPOGs ($10^{-1.21}$; \citealt{martinez-badenes+12,lisenfeld+14}). 

It is unclear which \lcomhtwo\ conversion factor is more relevant to our population of CO-SPOGs. \citet{a14_irtz,a16_sample} suggested that SPOGs were at an earlier stage of quenching than other poststarburst galaxies, and thus could still contain larger reservoirs of molecular gas as they undergo quenching. Observations of denser gas tracers, such as CS and HCN will be necessary to better infer the mass of dense gas and better define the $X_{\rm CO}$ factor.

\subsection{Comparing the CO(1--0) and 22$\mu$m flux}
The well-known connection between mid-IR emission and star formation \citep{calzetti+07} is caused by hot dust re-radiating UV photons from the young stars. Therefore, it is likely that a relation exists between the 22$\mu$m and CO(1--0) fluxes, given that the 22$\mu$m emission traces the star formation and the CO(1--0) traces the star-forming fuel \citep{rosenberg+15}. The left-most panel of Figure~\ref{fig:fco22} puts this relation to the test by comparing the 22$\mu$m and CO(1--0) fluxes of different samples of galaxies, including star-forming galaxies \citep{lisenfeld+11,coldgass,bauermeister+13}, early-type galaxies \citep{young+11,alatalo+13,davis+14}, radio galaxies \citep{evans+05}, and poststarburst galaxies \citep{french+15}.  The 22$\mu$m emission was obtained through a cross-matching with the ALLWISE catalog \citep{wise}. In cases that objects were flagged as extended in the 2-Micron All-Sky Survey (2MASS; \citealt{2mass}), the extended source flux was used ({\em w4gmag}), otherwise the profile fit flux ({\em w4mpro}) was used.

The star-forming objects of the COLD GASS survey \citep{coldgass}, Evolution of molecular Gas in Normal Galaxies (EGNoG; \citealt{bauermeister+13}), and Analysis of the interstellar Medium of Isolated GAlaxies (AMIGA;  \citealt{lisenfeld+11}) samples (black and gray dots) on Figure~\ref{fig:fco22} do indeed trace a reliable relation, allowing us to predict the CO(1--0) flux from the 22$\mu$m flux. We used the scaling of \citet{calzetti+07} of SFR\,$\propto$\,$L_{\rm 24\mu m}^{0.885}$, as well as \citet{carilli+walter13} of $L_{\rm FIR}$\,$\propto$\,$L_{\rm CO}^{1.4}$ (as well as $L_{\rm FIR}$ being linearly related to the SFR; \citealt{ken98}) to derive the expected slope of the relation between F(22\micron) and S$_{\rm CO}\Delta v$, setting it to: S$_{\rm CO}\Delta v$\,$\propto$\,F(22\micron)$^{0.632}$. Using a 4$\sigma$-clipped subset of the star-forming objects, and the relation noted above, we find the following relationship between $S_{\rm CO}\Delta v$ and $F_{\rm{22\mu}m}$:

\begin{equation}
\frac{S_{\rm CO}\Delta v}{\rm (Jy~km~s^{-1})} = 2.402^{+0.128}_{-0.122} \,\left(\frac{F_{\rm{22\mu}m}}{\rm mJy}\right)^{0.623}
\end{equation}

\noindent which corresponds to the bisecting line shown in Figure~\ref{fig:fco22}. For regular star-forming galaxies, this relation can predict a rough CO(1--0) flux, though the non-star forming samples show that this breaks down with the presence of other dominant contributors to the 22$\mu$m emission. 

The early-type galaxies from the ATLAS$^{\rm 3D}$ sample show a much larger scatter than the star-forming galaxies on this relation, also noted by \citet{davis+14}, with a slight mid-IR enhancement ($\epsilon_{\rm MIR}$; defined as the ratio of the 22$\mu$m emission in the source to the expected 22$\mu$m at a given $S_{\rm CO}\Delta v$ based on the star-forming objects defined line), $\langle\epsilon_{\rm MIR, ETG}\rangle$\,=\,4.22$^{+0.90}_{-0.74}$. This scatter is likely due to the fact that there are other sources of 22$\mu$m emission in early-type galaxies, including the aged stellar population and the hard ultraviolet field created by post-AGB stars (responsible for the UV upturn phenomenon; \citealt{oconnell99,davis+14}).

A subset of the radio galaxies and quasars \citep{evans+05} tend to diverge the most from this relation, as their 22$\mu$m emission over-predicts the CO(1--0) flux with $\langle\epsilon_{\rm MIR, AGN}\rangle$\,=\,5.58$^{+2.53}_{-1.75}$. Many of the galaxies that sit near the relation are known starbursting AGN hosts \citep{lanz+16}. The enhancement therefore is unsurprising, given that AGNs contribute substantially to the hot dust component of a galaxy's spectral energy distribution (SED) in the mid-infrared \citep{ward+87,sanders+89,elvis+94}. Thus, a quasar's divergence from this relation is due to the fact that star formation is not the primary contributor to the 22$\mu$m emission.

The poststarburst galaxies from \citet{french+15} appear to be strongly MIR-enhanced, sitting below the $S_{\rm CO}\Delta v$--$F_{\rm{22\mu}m}$ relation by a factor of $\langle\epsilon_{\rm MIR, PSB}\rangle$\,=\,10.51$^{+1.89}_{-1.60}$. An additional heating mechanism (that would only heat the dust, but not increase the CO flux by injecting turbulence) must be present in poststarburst galaxies to cause this enhancement, such as the radiation from the intermediate-aged stellar population or deeply buried AGNs.

CO-SPOGs are also a divergent population, albeit not to the extent of the poststarbursts, with all objects lying below the $S_{\rm CO}\Delta v$--$F_{\rm{22\mu}m}$ relation set by the star-forming galaxies with $\langle\epsilon_{\rm MIR, SPOG}\rangle$\,=\,4.91$^{+0.42}_{-0.39}$. In fact, if we limit ourselves only to strong (S/N$\geq$5) detections, $\langle\epsilon_{\rm MIR, SPOG}\rangle$ increases to 9.75$^{+1.91}_{-1.60}$. This suggests that there could be AGNs in our CO-SPOGs, but that the bolometric luminosities of these AGNs are not as strong as the extreme objects in the \citet{evans+05} sample. The selection criteria we applied to create the CO-SPOG sample (intermediate aged stars, a lack of star formation from the line diagnostic diagram, and detectable 22$\mu$m emission) favor AGNs; thus, the $S_{\rm CO}\Delta v$--$F_{\rm{22\mu}m}$ relation indicating the presence of AGNs is expected. How CO-SPOGs and poststarbursts compare will be presented in detail in \S\ref{sec:postsb}.

Their positions in Figure~\ref{fig:fco22} predict that SPOGs have AGN luminosities that are intermediate between the star-forming population and the quasar/radio galaxy population. Given the prevalence of intermediate-aged stars found in quasars \citep{canalizo+13}, it is possible that some SPOGs in our sample may migrate to more luminous AGN phases as the galaxy transition proceeds. 

The right panels of Figure~\ref{fig:fco22} have taken the relation and broken CO-SPOGs down into sub-populations based on their \nad\ properties (top), radio detections (middle), and morphological properties (bottom) to determine whether any of these properties influence $\epsilon_{\rm MIR}$. Our CO-SPOGs have $\langle\epsilon_{\rm MIR, SPOG}\rangle$\,=\,4.91$^{+0.42}_{-0.39}$, falling off the star forming relation by a factor of $\approx$\,5. Objects with either radio emission or a \nad\ enhancement (detailed in \S\ref{sec:nad}) appear to fall slightly farther from the relation with $\langle\epsilon_{\rm MIR, radio}\rangle$\,=\,4.95$^{+0.62}_{-0.55}$ and $\langle\epsilon_{\rm MIR, Na\,I\,D}\rangle$\,=\,5.40$^{+0.85}_{-0.73}$, respectively, but agree within the uncertainty. CO-SPOGs that were classified as morphologically disrupted were found to have smaller deviations from the relation, $\langle\epsilon_{\rm MIR, disrupted}\rangle$\,=\,4.37$^{+0.47}_{-0.43}$.  In all of these cases, the overall $\epsilon_{\rm MIR}$ was not sufficiently deviant from the overall CO-SPOG population to be distinct, and thus we cannot conclude whether the presence of any of these properties enhance or suppress the mid-IR emission as compared to the CO flux (relative to star forming galaxies).

\subsection{Ambiguity of star formation tracers in CO-SPOGs}
In CO-SPOGs, shocks can dominate the ionized gas emission \citep{allen+08,rich+11}, including [O\,{\sc ii}] and H$\alpha$. UV emission suffers from large uncertainties due to extinction and contamination due to non-star formation dominated phenomena, such as heating by intermediate-age stars. Finally, as Figure~\ref{fig:fco22} laid out, the 22$\mu$m emission can be significantly contaminated by the presence of an AGN. In NGC\,1266, the 22$\mu$m emission overestimates the SFR by a factor of 10 \citep{alatalo+11}, despite star formation being suppressed by a factor of $>$\,50 \citep{a15_sfsupp}. While centimeter-wave radio emission is a sensitive tracer of recent star formation \citep{condon92} the continuum emission at 1.4\,GHz can also be contaminated by the synchrotron emission from an AGN \citep{zensus+97,laor+08}.

Constructing a SED that includes far-infrared emission is by far the most reliable tracer of star formation in all systems except those with the most deeply buried AGNs, because the cold dust nearly unambiguously traces star formation originating in imbedded clouds. It is possible that a single data point on the Rayleigh-Jeans tail of the dust continuum blackbody could reduce the uncertainty of SFRs in ambiguous systems, such as SPOGs. For instance, an observation at 850$\mu$m would be able to anchor the blackbody. If we assume that the dust temperatures do not vary much away from $T_d$\,$\approx$\,25K \citep{scoville+14}, we should be able to use the 22$\mu$m and 850$\mu$m points to interpolate the modified blackbody for the dust continuum across, inferring a star formation rate. While this method is not as precise as fitting a full SED and dust continuum (with far-IR data near the peak of the blackbody), it could significantly improve SFR estimates for SPOGs.

\subsection{CO-SPOGs and CO-detected poststarbursts}
\label{sec:postsb}

\begin{figure}[t]
\subfigure{\includegraphics[width=0.49\textwidth]{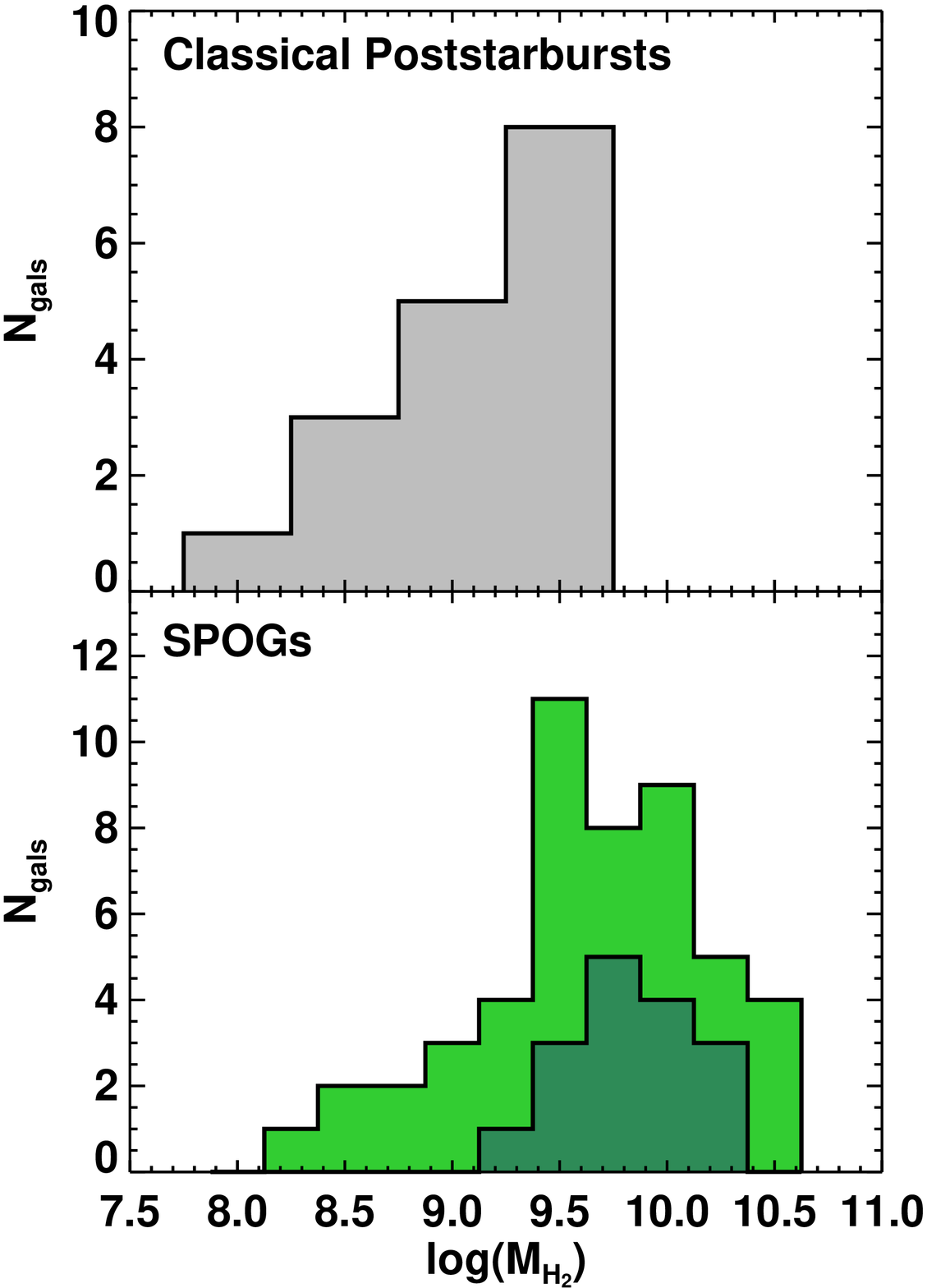}}
\caption{The distribution of molecular gas in the classical poststarburst galaxies (top; \citealt{french+15}) and CO-SPOGs (bottom, light green).  The distribution for the morphologically disrupted SPOGs is overlaid in dark green, and reflects the overall SPOG distribution faithfully. SPOGs have significantly more molecular mass than poststarbursts.}
\label{fig:histmh2}
\end{figure}

\citet{a16_sample}  argue that the SPOGs criterion might identify objects at an earlier stage of transition than the classical poststarburst criterion. \citet{french+15} showed that 53\% of poststarburst galaxies contain non-negligible reservoirs of molecular gas, modifying the standard picture of galaxy evolution in which a galaxy expels its interstellar medium before transitioning \citep{hopkins+06}. We have also shown that compared to their molecular gas content, the poststarbursts of \citet{french+15} are significantly 22$\mu$m-enhanced. 

The \citet{french+15} poststarburst sample has reservoirs of molecular gas ranging from 10$^{8.6}$--10$^{9.8}$\,$M_\odot$ in mass, and molecular-to-stellar mass fractions between 10$^{-2}$--10$^{-0.5}$. In that sample, the majority of the poststarburst galaxies contain ionized gas line ratios that are consistent with LINERs, and a substantial fraction show disrupted optical morphologies. Of the 17 CO-detected poststarbursts from the \citet{french+15} sample, 15 were detected in the 22$\mu$m WISE band with a S/N ratio $>$3, thus all but 2 \citet{french+15} poststarbursts would have surpassed the 22$\mu$m criterion of the CO-SPOGs. There is no difference in average $f_{\rm mol}$ in the poststarburst sample when we remove the two 22$\mu$m non-detected postsarbursts from consideration.

CO-SPOGs compare well to the \citet{french+15} poststarbursts in ionized gas ratios and 22$\mu$m properties, and morphologies, including both sets presenting substantial LINER-like ionized gas line diagnostics, similar detection rates in {\em WISE} 22$\mu$m emission, and a similar fraction of objects present morphological disruptions. Due to a selection against strong [O\,{\sc iii}] emission, the \citet{french+15} CO-detected poststarbursts have an average EW[O\,{\sc iii}] that is an order of magnitude smaller ($\langle$EW[O\,{\sc iii}]$_{\rm PSB}\rangle$\,=\,0.60) than what is found for CO-SPOGs ($\langle$EW[O\,{\sc iii}]$_{\rm COSPOG}\rangle$\,=\,5.74). The poststarbursts of \citet{french+15} show low CO fluxes compared to their 22$\mu$m fluxes, and CO-SPOGs seem to contain larger reservoirs of molecular gas (Fig.~\ref{fig:histmh2}; assuming equivalent values for $X_{\rm CO}$). We also see discrepancies between SPOGs and poststarbursts  in the molecular-to-stellar mass ratio (Figure~\ref{fig:fmol}), where SPOGs exhibit much higher gas fractions.\footnote{A standard $X_{\rm CO}$ was used both for poststarbursts and for CO-SPOGs, since the time differentiation between the SPOG phase and poststarburst phase is sufficiently short that we do not expect a dramatic change in the conversion outside of the scatter seen in different merger simulations \citep{narayanan+11}.}

The fact that SPOGs contain more molecular gas than \citet{french+15} poststarbursts could suggest one of the following. Either (1) the SPOG criteria allowed for galaxies that contained more substantial reservoirs of molecular gas in the first place, due to not strictly excluding H$\alpha$ or [O\,{\sc ii}] emission, or (2) SPOGs are in an earlier phase of their transition; possibly both of these effects are at play. It is also possible that in some CO-SPOGs, the bulk of star formation is taking place behind an optically thick screen, although IFU studies of objects at different merger stages have shown that such ``skin effect'' phenomena are rare \citep{rich+15} and do not explain the excess 22\micron\ enhancement present in the poststarbursts. Optical IFU studies of CO-SPOGs would help elucidate whether this effect is present in any of our objects.

Figure~\textcolor{Maroon}{9} in \citet{a16_sample} indicates that SPOGs are bluer than poststarburst galaxies \citep{goto07,yesuf+14,rowlands+15} and lie on the blueward side of the green valley, and Figure~\ref{fig:co_cmd} confirms that CO-SPOGs follow the colors of the general SPOG population.  SPOG optical colors may be bluer than poststarbursts because they have truncated their star formation more recently, and thus have a larger population of blue stars (to contribute to the galaxy colors). A large-scale stellar population analysis of these two populations would be able to better pinpoint how SPOGs and poststarbursts are related, but the larger reservoir of molecular gas in SPOGs could suggest that they are at an earlier phase of evolution, if depletion of the molecular reservoir is related to the time since the start of a galaxy's transition.

If galaxy transition does correlate with the depletion of molecular material, then the larger gas reservoirs and gas fractions are suggestive of a timescale effect, which correlates with the phase of transition. \citet{a15_hcgco} used interferometric CO measurements to show that in some warm H$_2$-bright HCG galaxies, the expulsion of the molecular reservoir was not required for a galaxy to transform. Instead, these authors argue that the inhibition of the molecular gas to form stars was a more important factor in predicting a galaxy's color. Whether CO-SPOGs or the poststarbursts of \citet{french+15} show this impact too will require sufficiently resolved molecular gas and accurate star formation rates to study whether a star formation suppression phase globally manifests in transforming galaxies or whether it is a special phenomenon seen in the compact group environment.

\begin{figure}[t]
\subfigure{\includegraphics[width=0.47\textwidth]{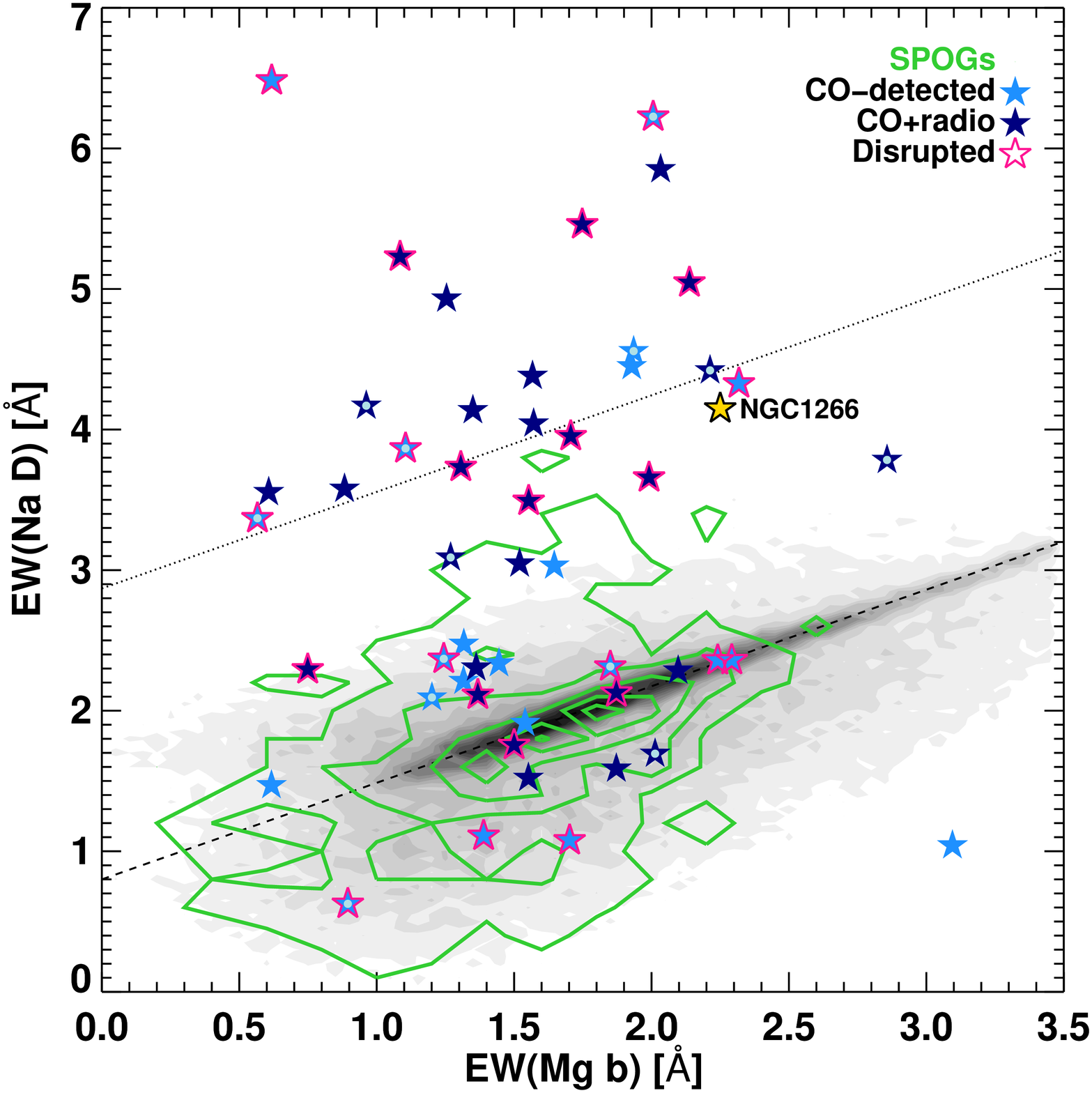}}
\caption{The equivalent widths EW(Mg\,b) vs. EW(\nad) of the ELG sample (grayscale; \citealt{a16_sample}) compared with the entire SPOGS sample (green contours). The dashed line represents the empirical relation found in \citet{a16_sample}. The dotted line represents a 3$\sigma$ departure from the empirical relation. The CO-SPOG sample is shown (stars), including FIRST radio detections (dark blue) and non-detections (light blue), as well as NGC\,1266 (gold; \citealt{davis+12}). CO-SPOGs with a light blue dot in their centers are detected with S/N between 3--5. SPOGs have \nad\ compared to Mg\,b that is enhanced beyond what is seen in normal star-forming galaxies in the ELG sample. The CO(1--0) detected objects show even more enhanced \nad\ characteristics compared to non-CO observed SPOGs, with 19 (37\%) objects beyond the 3$\sigma$ boundary defined by the normal (ELG-defined) relation. There does not appear to be a difference between the objects observed by CARMA and those observed by the IRAM~30m. Clearly and possibly morphologically disrupted SPOGs are outlined in pink. Radio-detected SPOGs tend to have the largest \nad\ excess, and radio non-detected, non-disrupted SPOGs have the least excess. Most radio non-detected SPOGs that do show \nad\ excess at the $>$3$\sigma$ level also show morphological disruptions.}
\label{fig:NaD}
\end{figure}

\subsection{\nad\ in CO-SPOGs: do they contain neutral winds?}
\label{sec:nad}

\citet{a16_sample} showed that, as a population, SPOGs have a higher fraction of sources with \nad\ properties that require an interstellar component compared to the (star formation-dominated) ELG sample. Figure~\ref{fig:NaD} shows the underlying distribution from the ELG sample (in grayscale), along with the SPOG distribution (green contours), where we see that there is a \nad-enhanced wing. Overplotted are CO-SPOGs (stars), labeled based on whether they were detected at 1.4\,GHz by FIRST (dark blue) or not (light blue), and whether they are disrupted (magenta outline). We can see that the CO-SPOGs are even more \nad-enhanced than the underlying SPOG population. We ran the Mann Whitney U-test to test whether the \nad\ properties of CO-SPOGs could be randomly drawn from the SPOG sample, and found that the null hypothesis was ruled out ({\em p}\,$\approx$\,0). 19 (37$\pm$7\%) of the CO-SPOGs contain \nad\ emission above the 3$\sigma$ boundary defined by the \nad-Mg\,b relation of the ELG sample (Equation 10 in \citealt{a16_sample}).  This excess suggests the possibility that many CO-SPOGs host interstellar winds \citep{rupke+05,veilleux+05,murray+07,park+15}.  

Figure~\ref{fig:NaD} indicates that a fraction of CO-SPOGs host \nad\ properties as extreme as what is seen in AGN-driven outflow hosts \citep{rupke+11,rupke+15}, including NGC\,1266 \citep{alatalo+11,davis+12,nyland+13}. The extreme \nad\ widths of these sources suggests that they may host an AGN-driven molecular outflow like NGC\,1266. If so, it would indeed shed light on how such outflows relate to galaxy transformation. However, it is also possible that \nad\ enhancements are due to unsettled neutral gas along the line of sight instead.

We have investigated how \nad\ absorption changes across the ELG sample.  Table~\ref{tab:nad} shows the median values, and 10$^{\rm th}$ and 90$^{\rm th}$ percentile values of the ELG, SPOG, and CO-SPOG subsamples.\footnote{It is of note that there is very little difference in the \nad\ properties of strong (S/N$\geq$5) detections and tentative (3$\leq$S/N$<$5) detections.} Overall, objects that have been detected in the 22$\mu$m band of {\em WISE} appear to show more enhanced \nad\ in the both ELG and the SPOG samples. In the case of galaxies within the ELG (and thus star formation dominant) sample, this \nad\ enhancement might be able to be explained by neutral winds being launched by the starburst \citep{murray+07,park+15,sarzi+16}, which would correlate with the 22$\mu$m hot dust emission (and therefore, the star formation rate).  The SPOG selection criteria eliminated strong star-formers (based on ionized gas emission line ratios; \citealt{bpt,veilleux+87,kewley+06}), so the \nad\ excess and 22$\mu$m emission are possibly from an alternative source. 

As discussed in \S\ref{sec:sample}, the ramifications of this selection are that our CO-SPOGs likely favor the presence of AGNs. SPOGs also follow the trend that objects hosting radio emission and 22$\mu$m emission show a significant \nad\ enhancement when compared to other samples, including radio or 22$\mu$m non-detected SPOGs. Radio-detected SPOGs exhibit a median \nad-enhancement (defined as the deviation from the mean relation of the ELG of \nad\,=\,(0.685\,Mg\,b+0.8) from \citealt{a16_sample}) of $\langle\epsilon_{\rm NaD, radio}\rangle$\,=\,1.421, and 13/19 (68$\pm$11\%) objects found with 3$\sigma$ \nad\ enhancements were radio-detected (although 57$\pm$7\% of the CO-SPOGs have been detected in radio). Objects that were not detected in FIRST have a median \nad\ enhancement of $\langle\epsilon_{\rm NaD,non-radio}\rangle$\,=\,0.547. CO-SPOGs with radio emission have higher \nad\ enhancements than objects that do not. This is further supported by the results of the Mann-Whitney U-test, which was able to rule out that the \nad\ properties of radio non-detected CO-SPOGs were drawn from the same distribution of radio-detected CO-SPOGs with the probability of a null hypothesis of {\em p}\,$<$\,0.04. Given that AGN activity is a probable origin of the radio emission, we suggest that AGN-hosting SPOGs are the most likely to contain enhanced \nad.

\begin{table}[t!]
\centering
\caption{\nad\ properties of SPOGs} \vspace{1mm}
\begin{tabular}{l r r r r}
\hline \hline
{\bf Type} & N$_{\rm obj}$& $\mathbf{\widetilde{\epsilon_{\rm NaD}}}$ & $\epsilon_{\rm NaD, 10}$ & $\epsilon_{\rm NaD, 90}$\\
& (1) & (2) & (3) & (4) \\
\hline
ELG & 159,387 & -0.073 & -1.142 & 0.833\\
ELG 22\micron$>$3$\sigma$& 71,301 & -0.009 & -0.966 & 1.023\\
\hline
SPOGs & 1,067 & -0.043 & -1.111 & 1.656\\
SPOG 22\micron$>$3$\sigma$ & 491 & 0.115 & -0.856 & 2.105\\
\hline
CO-SPOGs & 52 & 1.421 & -0.485 & 3.273\\
CO-SPOGs, S/N$>$5 & 34 & 1.452 & -0.341 & 3.463\\
CO-SPOGs, radio & 30 & 1.984 & -0.072 & 3.463\\
CO-SPOGs, morph. & 24 & 1.495 & -0.641 & 3.686\\
\hline \hline
\end{tabular} \\
\label{tab:nad}
\raggedright {\footnotesize
{Column (1): Total number of objects in each class. Column (2): Median \nad\ enhancement of each sample. Column (3): The 10$^{\rm th}$ percentile \nad\ enhancement. Column (4): The 90$^{\rm th}$ percentile \nad\ enhancement.}
}
\vspace{1mm}
\end{table}

We also tested whether the morphologies of the SPOGs had a strong influence on the \nad\ enhancement, given that many objects that host strong neutral winds are starbursts in ULIRGs, which are mostly major mergers \citep{veilleux+05}. The overall effect of morphological disruption seems to have a slight impact on the \nad\ enhancement, with median $\langle\epsilon_{\rm NaD, disrupted}\rangle$\,=\,1.495, slightly larger than the median for galaxies that were not classified as being disrupted of $\langle\epsilon_{\rm NaD, undisturbed}\rangle$\,=\,1.206. Therefore, morphological disruption appears to have a smaller influence on the \nad\ enhancement of CO-SPOGs than the presence of radio emission.

\citet{sarzi+16} showed that in a sample of 456 nearby galaxies, sources with \nad\ enhancements attributable to interstellar winds (as opposed to [Na/Fe] overabundances as is seen in the most massive galaxies; \citealt{jeong+13}) were not observed in objects detected with Very Long Baseline Interferometry \citep{mjive}. These authors concluded that the majority of \nad-enhanced objects were prolifically star-forming galaxies with neutral winds that were star formation driven, and that \nad\ winds originate most commonly in star-forming galaxies due to star formation driving. In CO-SPOGs, it appears that there are (radio-detected) AGNs and \nad\ enhancements in the same objects, consistent with the results of \citet{lehnert+11} that 33\% of radio galaxies within SDSS also contain \nad\ enhancements. The lack of this type of object within the \citet{sarzi+16} sample suggests that they could be quite rare or are possibly a short and violent phase in galaxy evolution, in which the star formation has quenched and an AGN is driving a neutral wind that is stirring up the remaining interstellar medium. The SPOG criterion selected against star-forming objects (which is able to rule out strong starbursts that are needed to drive winds; \citealt{a16_sample}), and thus include many more galaxies with LINER emission, as opposed to the \citet{sarzi+16} sample.

Further studies are needed, including observations with higher spectral resolution to investigate the shape and structure of the \nad\ line, to look for outflows \citep{veilleux+05,rupke+05,rupke+11}. Integral field spectrographs will be able to determine the extent of the \nad\ absorption, provide high resolution kinematics to determine the neutral mass flux and compare the \nad\ kinematics to the stellar kinematics. Deeper CO observations might detect broad molecular wings, resulting in the measurement of a range of the mass outflow (and mass escape) rates observed in transitioning objects. In-depth studies of these objects' star formation rates and molecular gas distributions will shed light on how often star formation is suppressed, leading to the conserving and extending the lifetime of the molecular gas as galaxies undergo transformations from late-types to early-types. Deep X-ray observations will put limits on the AGN luminosity, helping to determine the range in energy budgets these systems might exhibit, and 2-dimensional stellar population studies might be able to provide a range of timescales over which triggering mechanisms started the process of star formation quenching in these systems.

\section{Summary}
\label{sec:summary}
We have followed up 52 of the {\em WISE} 22$\mu$m-detected objects from the Shocked POststarburst Galaxy Survey using the IRAM~30m and CARMA to search for CO(1--0). We were able to detect 47 of these 52 SPOGs to at least 3$\sigma$ significance. 

\begin{itemize} 
\item The requirement of detected 22$\mu$m emission, combined with ionized gas emission line ratios {\em inconsistent} with star formation, likely biases our CO-SPOG sample toward the detection of AGNs. Despite this, a large subset of our objects do not have line ratios consistent with a pure Seyfert nucleus.

\item The CO-SPOG sample appears to span the color phase space of the SPOG parent sample, though with a bias toward the more massive SPOGs. A morphological investigation was undertaken to visually classify whether an object was disrupted, finding that 37--46$\pm$7\% of our CO-SPOG sample show signs of morphological disruptions.

\item The molecular gas fractions exhibited by the CO-SPOGs are larger than those in normal star-forming galaxies and those in a sample of traditionally-identified poststarburst galaxies, most of which were detected in 22$\mu$m emission. The molecular gas fractions identified in our sample are consistent with those seen in interactions, supported by our identification of a large fraction of morphologically disrupted objects, although it is possible that our 22$\mu$m selection has biased the sample to select objects with buried ongoing star formation, which will require further observations to measure accurately.

\item We used star-forming galaxies to derive a relation between the 22$\mu$m flux from {\em WISE} and the CO(1--0) flux, finding that they were in agreement (supporting the claim that the mid-IR in star-forming galaxies is originating from the star formation itself) and that quasars and radio galaxies fall off this relation. SPOGs in general sit between star-forming galaxies and quasars/radio galaxies, with an average mid-IR enhancement of $\langle\epsilon_{\rm MIR}\rangle$\,=\,4.91$^{+0.42}_{-0.39}$. Presence of radio emission, \nad\ enhancement, or morphological disruption might influence $\epsilon_{\rm MIR}$, but not in a way that significantly deviates from the underlying CO-SPOG population.

\item The enhancement in the \nad\ absorption relative to the Mg\,b absorption is more significant in the CO-SPOGs than in the general SPOG population, with 19/52 (37$\pm$7\%) detected 3$\sigma$ above the  empirical relation from the original ELG sample. This may be due to the likely AGN over-population within the CO-SPOG sample, further supported by the larger \nad\ enhancement present in the radio-detected objects.


\end{itemize}

\acknowledgments
KA thanks K. Decker French for useful discussions regarding the ``E+A'' sample. KA also thanks Jonathan McDowell \& Michael J.\,I. Brown for Twitter dialogues regarding {\em WISE} and mid-infrared dust emission in AGNs, improving the manuscript. We also thank the anonymous referee for an insightful report that has improved the manuscript.

Support for KA is provided by NASA through Hubble Fellowship grant \hbox{\#HST-HF2-51352.001} awarded by the Space Telescope Science Institute, which is operated by the Association of Universities for Research in Astronomy, Inc., for NASA, under contract NAS5-26555. UL acknowledges support by the research projects AYA2011- 24728 and AYA2014-53506-P financed by the Spanish Ministerio de Econom\'ia y Competividad and by FEDER (Fondo Europeo de Desarrollo Regional) and the Junta de Andaluc\'\i a (Spain) grants FQM108. PNA is partially supported by funding through {\em Herschel}, a European Space Agency Cornerstone Mission with significant participation by NASA, through an award issued by JPL/Caltech. SLC was supported by ALMA-CONICYT program 31110020. KN acknowledges support from NASA through the {\em Spitzer} Space Telescope. AMM and LJK acknowledge the support of the Australian Research Council (ARC) through Discovery project DP130103925.

Based on observations carried out with the IRAM~30m Telescope. IRAM is supported by INSU/CNRS (France), MPG (Germany) and IGN (Spain).  Support for CARMA construction was derived from the Gordon and Betty Moore Foundation, the Kenneth T. and Eileen L. Norris Foundation, the James S. McDonnell Foundation, the Associates of the California Institute of Technology, the University of Chicago, the states of California, Illinois, and Maryland, and the National Science Foundation. Ongoing CARMA development and operations are supported by the National Science Foundation under a cooperative agreement, and by the CARMA partner universities. This publication makes use of data products from the {\em Wide-field Infrared Survey Explorer}, which is a joint project of the University of California, Los Angeles, and the Jet Propulsion Laboratory/California Institute of Technology, funded by the National Aeronautics and Space Administration. This research has made use of the NASA/ IPAC Infrared Science Archive, which is operated by the Jet Propulsion Laboratory, California Institute of Technology, under contract with the National Aeronautics and Space Administration. The National Radio Astronomy Observatory is a facility of the National Science Foundation operated under cooperative agreement by Associated Universities, Inc.\\

\noindent{\em Facilities:} \facility{CARMA}, \facility{IRAM}, \facility{{\em WISE}}

\appendix
\section{Gaussian fitting the IRAM 30m observations}
\label{app:gauss}
\begin{figure*}[h]
\centering
\includegraphics[width=0.97\textwidth]{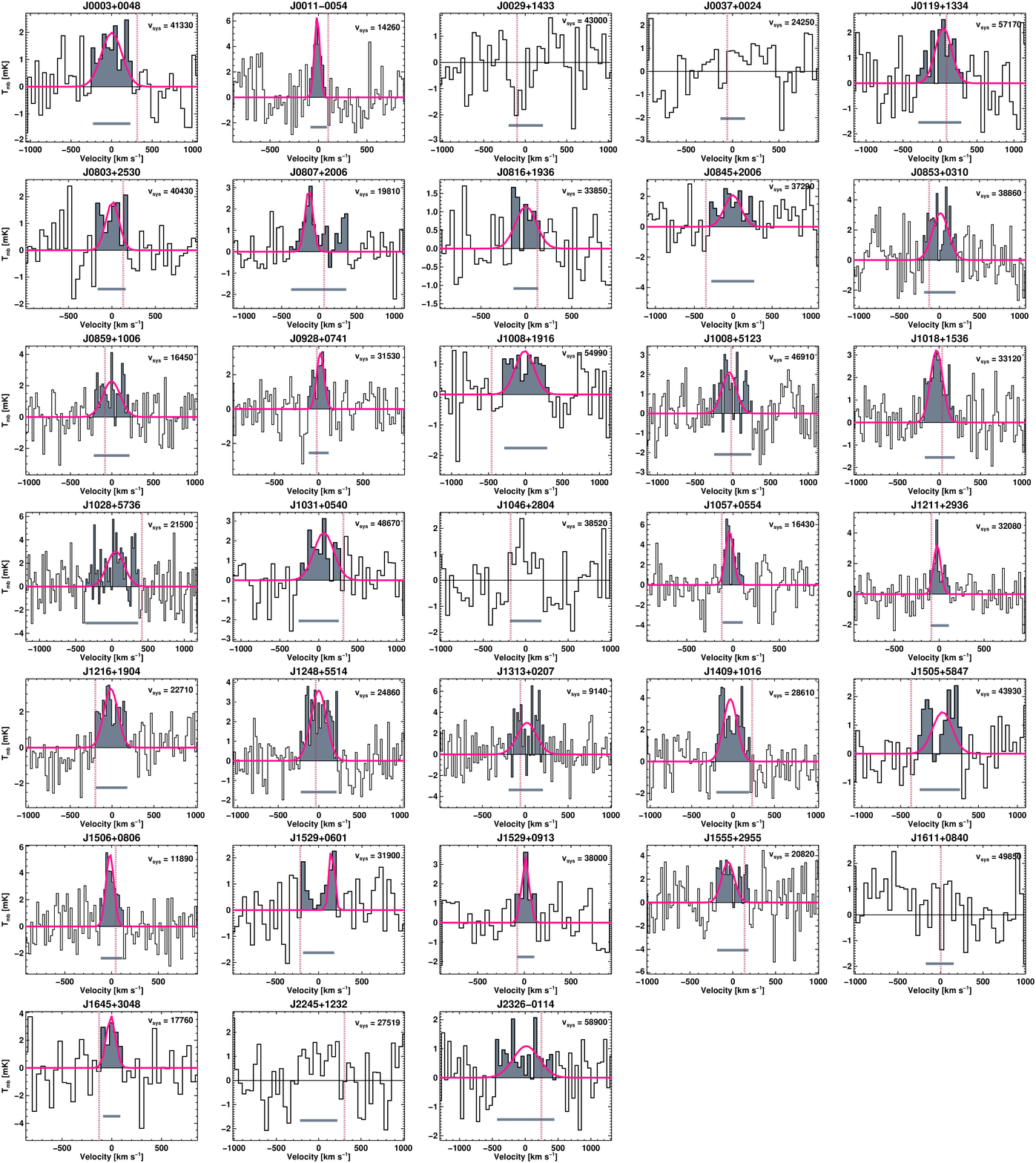}
\caption{The 33 IRAM~30m spectra (as in Fig.~\ref{fig:specs}), with the chosen linewidths shaded gray, overlaid with the best fit single Gaussian profile (pink), overlaid with the optically-determined velocity (dotted maroon line).}
\label{fig:gauss_iram}
\end{figure*}

\begin{figure}[h!]
\centering
\includegraphics[width=0.48\textwidth]{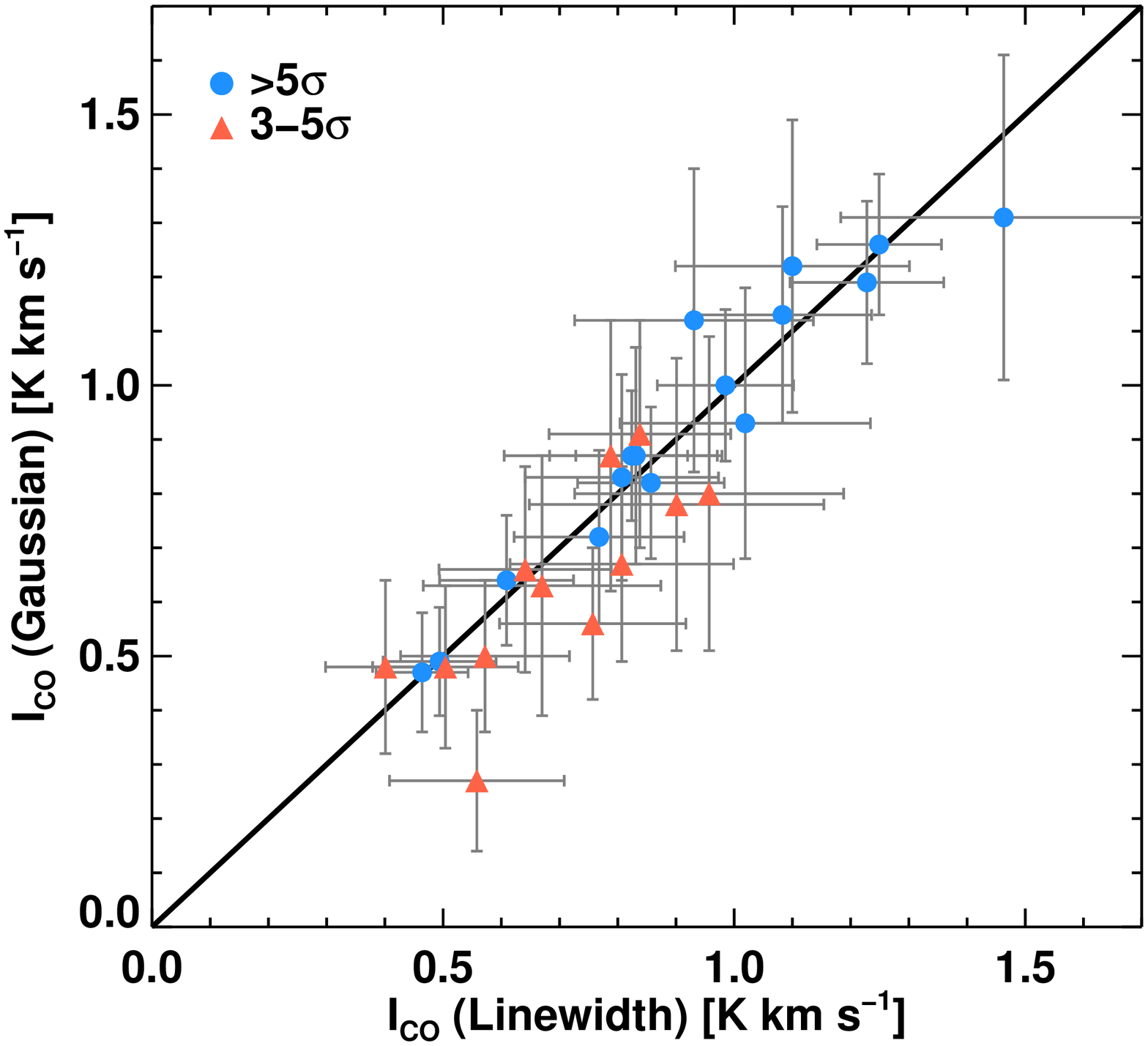}
\caption{A comparison between $I_{\rm CO}$ determined in both of the methods: fitting a single Gaussian vs. integrating the total intensity of the line for detected IRAM CO-SPOGs. Strong (S/N\,$>$5) detections are shown as blue circles and tentative (S/N\,=\,3--5) are shown as red triangles. The average disagreement between the two methods is 5\%, and the vast majority of objects agree within errors.}
\label{fig:gauss_comp}
\end{figure}

As a check of the accuracy of the fluxes measured by the IRAM~30m (especially those that with S/N ratios between 3--5), we fit each spectrum to a single Gaussian profile in order to investigate the likelihood that our detections are real or spurious. We used the {\tt minimize} procedure in {\sc gildas}\footnote{http://www.iram.fr/IRAMFR/GILDAS} on the continuum-subtracted, calibrated IRAM SPOG datasets (excluding non-detections). {\tt minimize}  was free to search for the line in a 3000\,km~s$^{-1}$ width centered on the recession velocity, but no initial velocity guess was provided for the CO(1--0) line. Figure~\ref{fig:gauss_iram} shows the results of the single Gaussian fits (pink) overlaid on the spectra of the IRAM SPOGs, with the chosen velocity widths shaded gray underlaid. The cases with a disagreement between the optical and radio recession velocities are most likely due to uncertainties in the SDSS recession velocities (individual SDSS spectral channels have widths of $\approx$\,100km~s$^{-1}$; \citealt{bolton+12}), although it is possible that the molecular gas in some of these systems is systemically offset from the stars and possibly part of a tidal tail or some disrupted gas structure, though consider the former possibility more likely. Overall, we can see that for the majority of sources, the Gaussian fitting faithfully detected and fit the profiles, although in some cases the SPOGs would have been better fit with an additional Gaussian.

Figure~\ref{fig:gauss_comp} shows the direct comparison between the intensities derived from both methods for strong detections ($>$5$\sigma$, blue points) and tentative detections (3--5$\sigma$, red triangles) within the IRAM sample. In all but two cases, the flux determinations match within errors, with an average mismatch of 5\%. In the cases where the Gaussian method has underestimated the total intensities, often at least 2 Gaussian profiles were needed for the fit (e.g., J1505+5847 and J1529+0601).

Overall, the Gaussian fits fluxes agreed well with our linewidth-determined fluxes, both in confirming the presence of the line at the optically-defined velocity, as well as in the total amount of flux that was detected.

\bibliographystyle{aasjournal}
\bibliography{../../../master}

\begin{thebibliography}{}
\expandafter\ifx\csname natexlab\endcsname\relax\def\natexlab#1{#1}\fi

\bibitem[{{Aalto} {et~al.}(1995){Aalto}, {Booth}, {Black}, \&
  {Johansson}}]{aalto+95}
{Aalto}, S., {Booth}, R.~S., {Black}, J.~H., \& {Johansson}, L.~E.~B. 1995,
  \aap, 300, 369

\bibitem[{{Aalto} {et~al.}(2016){Aalto}, {Costagliola}, {Muller},
  {et~al.}}]{aalto+16}
{Aalto}, S., {Costagliola}, F., {Muller}, S., {et~al.} 2016, \aap, 590, A73

\bibitem[{{Aalto} {et~al.}(2012{\natexlab{a}}){Aalto}, {Garcia-Burillo},
  {Muller}, {et~al.}}]{aalto+12}
{Aalto}, S., {Garcia-Burillo}, S., {Muller}, S., {et~al.} 2012{\natexlab{a}},
  \aap, 537, A44

\bibitem[{{Aalto} {et~al.}(2012{\natexlab{b}}){Aalto}, {Muller}, {Sakamoto},
  {et~al.}}]{aalto_1377}
{Aalto}, S., {Muller}, S., {Sakamoto}, K., {et~al.} 2012{\natexlab{b}}, \aap,
  546, A68

\bibitem[{{Alatalo}(2015)}]{alatalo15}
{Alatalo}, K. 2015, \apjl, 801, L17

\bibitem[{{Alatalo} {et~al.}(2015{\natexlab{a}}){Alatalo}, {Appleton},
  {Lisenfeld}, {et~al.}}]{a15_hcgco}
{Alatalo}, K., {Appleton}, P.~N., {Lisenfeld}, U., {et~al.} 2015{\natexlab{a}},
  \apj, 812, 117

\bibitem[{{Alatalo} {et~al.}(2011){Alatalo}, {Blitz}, {Young},
  {et~al.}}]{alatalo+11}
{Alatalo}, K., {Blitz}, L., {Young}, L.~M., {et~al.} 2011, \apj, 735, 88

\bibitem[{{Alatalo} {et~al.}(2014{\natexlab{a}}){Alatalo}, {Cales}, {Appleton},
  {et~al.}}]{a14_irtz}
{Alatalo}, K., {Cales}, S.~L., {Appleton}, P.~N., {et~al.} 2014{\natexlab{a}},
  \apjl, 794, L13

\bibitem[{{Alatalo} {et~al.}(2016){Alatalo}, {Cales}, {Rich},
  {et~al.}}]{a16_sample}
{Alatalo}, K., {Cales}, S.~L., {Rich}, J.~A., {et~al.} 2016, \apjs, 224, 38

\bibitem[{{Alatalo} {et~al.}(2015{\natexlab{b}}){Alatalo}, {Crocker}, {Aalto},
  {et~al.}}]{a15_co13}
{Alatalo}, K., {Crocker}, A.~F., {Aalto}, S., {et~al.} 2015{\natexlab{b}},
  \mnras, 450, 3874

\bibitem[{{Alatalo} {et~al.}(2013){Alatalo}, {Davis}, {Bureau},
  {et~al.}}]{alatalo+13}
{Alatalo}, K., {Davis}, T.~A., {Bureau}, M., {et~al.} 2013, \mnras, 432, 1796

\bibitem[{{Alatalo} {et~al.}(2015{\natexlab{c}}){Alatalo}, {Lacy}, {Lanz},
  {et~al.}}]{a15_sfsupp}
{Alatalo}, K., {Lacy}, M., {Lanz}, L., {et~al.} 2015{\natexlab{c}}, \apj, 798,
  31

\bibitem[{{Alatalo} {et~al.}(2014{\natexlab{b}}){Alatalo}, {Nyland}, {Graves},
  {et~al.}}]{a14_stelpop}
{Alatalo}, K., {Nyland}, K., {Graves}, G., {et~al.} 2014{\natexlab{b}}, \apj,
  780, 186

\bibitem[{{Allen} {et~al.}(2008){Allen}, {Groves}, {Dopita},
  {et~al.}}]{allen+08}
{Allen}, M.~G., {Groves}, B.~A., {Dopita}, M.~A., {et~al.} 2008, \apjs, 178, 20

\bibitem[{{Appleton} {et~al.}(2014){Appleton}, {Mundell}, {Bitsakis},
  {et~al.}}]{appleton+14}
{Appleton}, P.~N., {Mundell}, C., {Bitsakis}, T., {et~al.} 2014, \apj, 797, 117

\bibitem[{{Baade}(1958)}]{baade58}
{Baade}, W. 1958, Ricerche Astronomiche, 5, 3

\bibitem[{{Baldry} {et~al.}(2004){Baldry}, {Glazebrook}, {Brinkmann},
  {et~al.}}]{baldry+04}
{Baldry}, I.~K., {Glazebrook}, K., {Brinkmann}, J., {et~al.} 2004, \apj, 600,
  681

\bibitem[{{Baldwin} {et~al.}(1981){Baldwin}, {Phillips}, \& {Terlevich}}]{bpt}
{Baldwin}, J.~A., {Phillips}, M.~M., \& {Terlevich}, R. 1981, \pasp, 93, 5

\bibitem[{{Bauermeister} {et~al.}(2013){Bauermeister}, {Blitz}, {Bolatto},
  {et~al.}}]{bauermeister+13}
{Bauermeister}, A., {Blitz}, L., {Bolatto}, A., {et~al.} 2013, \apj, 768, 132

\bibitem[{{Becker} {et~al.}(1995){Becker}, {White}, \& {Helfand}}]{first}
{Becker}, R.~H., {White}, R.~L., \& {Helfand}, D.~J. 1995, \apj, 450, 559

\bibitem[{{Bekki} {et~al.}(2002){Bekki}, {Couch}, \& {Shioya}}]{bekki+02}
{Bekki}, K., {Couch}, W.~J., \& {Shioya}, Y. 2002, \apj, 577, 651

\bibitem[{{Bell} {et~al.}(2003){Bell}, {McIntosh}, {Katz}, \&
  {Weinberg}}]{bell+03}
{Bell}, E.~F., {McIntosh}, D.~H., {Katz}, N., \& {Weinberg}, M.~D. 2003, \apjs,
  149, 289

\bibitem[{{Bitsakis} {et~al.}(2014){Bitsakis}, {Charmandaris}, {Appleton},
  {et~al.}}]{bitsakis+14}
{Bitsakis}, T., {Charmandaris}, V., {Appleton}, P.~N., {et~al.} 2014, \aap,
  565, A25

\bibitem[{{Bitsakis} {et~al.}(2011){Bitsakis}, {Charmandaris}, {da Cunha},
  {et~al.}}]{bitsakis+11}
{Bitsakis}, T., {Charmandaris}, V., {da Cunha}, E., {et~al.} 2011, \aap, 533,
  A142

\bibitem[{{Bitsakis} {et~al.}(2010){Bitsakis}, {Charmandaris}, {Le Floc'h},
  {et~al.}}]{bitsakis+10}
{Bitsakis}, T., {Charmandaris}, V., {Le Floc'h}, E., {et~al.} 2010, \aap, 517,
  A75

\bibitem[{{Blanton} \& {Moustakas}(2009)}]{blanton+moustakas09}
{Blanton}, M.~R., \& {Moustakas}, J. 2009, \araa, 47, 159

\bibitem[{{Bock} {et~al.}(2006){Bock}, {Bolatto}, {Hawkins}, {et~al.}}]{carma}
{Bock}, D.~C.-J., {Bolatto}, A.~D., {Hawkins}, D.~W., {et~al.} 2006, in Society
  of Photo-Optical Instrumentation Engineers (SPIE) Conference Series, Vol.
  6267, Society of Photo-Optical Instrumentation Engineers (SPIE) Conference
  Series, 626713

\bibitem[{{Bolatto} {et~al.}(2013){Bolatto}, {Wolfire}, \&
  {Leroy}}]{bolatto+13}
{Bolatto}, A.~D., {Wolfire}, M., \& {Leroy}, A.~K. 2013, \araa, 51, 207

\bibitem[{{Bolton} {et~al.}(2012){Bolton}, {Schlegel}, {Aubourg},
  {et~al.}}]{bolton+12}
{Bolton}, A.~S., {Schlegel}, D.~J., {Aubourg}, {\'E}., {et~al.} 2012, \aj, 144,
  144

\bibitem[{{Cales} {et~al.}(2011){Cales}, {Brotherton}, {Shang},
  {et~al.}}]{cales+11}
{Cales}, S.~L., {Brotherton}, M.~S., {Shang}, Z., {et~al.} 2011, \apj, 741, 106

\bibitem[{{Cales} {et~al.}(2013){Cales}, {Brotherton}, {Shang},
  {et~al.}}]{cales+13}
---. 2013, \apj, 762, 90

\bibitem[{{Calzetti} {et~al.}(2007){Calzetti}, {Kennicutt}, {Engelbracht},
  {et~al.}}]{calzetti+07}
{Calzetti}, D., {Kennicutt}, R.~C., {Engelbracht}, C.~W., {et~al.} 2007, \apj,
  666, 870

\bibitem[{{Canalizo} \& {Stockton}(2013)}]{canalizo+13}
{Canalizo}, G., \& {Stockton}, A. 2013, \apj, 772, 132

\bibitem[{{Cannon} \& {Pickering}(1918)}]{obafgkm}
{Cannon}, A.~J., \& {Pickering}, E.~C. 1918, Annals of Harvard College
  Observatory, 91, 1

\bibitem[{{Carilli} \& {Walter}(2013)}]{carilli+walter13}
{Carilli}, C.~L., \& {Walter}, F. 2013, \araa, 51, 105

\bibitem[{{Carter} {et~al.}(2012){Carter}, {Lazareff}, {Maier},
  {et~al.}}]{emir}
{Carter}, M., {Lazareff}, B., {Maier}, D., {et~al.} 2012, \aap, 538, A89

\bibitem[{{Casey} {et~al.}(2014){Casey}, {Scoville}, {Sanders},
  {et~al.}}]{casey+14}
{Casey}, C.~M., {Scoville}, N.~Z., {Sanders}, D.~B., {et~al.} 2014, \apj, 796,
  95

\bibitem[{{Chung} {et~al.}(2009){Chung}, {van Gorkom}, {Kenney}, {Crowl}, \&
  {Vollmer}}]{chung+09b}
{Chung}, A., {van Gorkom}, J.~H., {Kenney}, J.~D.~P., {Crowl}, H., \&
  {Vollmer}, B. 2009, \aj, 138, 1741

\bibitem[{{Cicone} {et~al.}(2012){Cicone}, {Feruglio}, {Maiolino},
  {et~al.}}]{cicone+12}
{Cicone}, C., {Feruglio}, C., {Maiolino}, R., {et~al.} 2012, \aap, 543, A99

\bibitem[{{Cicone} {et~al.}(2014){Cicone}, {Maiolino}, {Sturm},
  {et~al.}}]{cicone+14}
{Cicone}, C., {Maiolino}, R., {Sturm}, E., {et~al.} 2014, \aap, 562, A21

\bibitem[{{Cluver} {et~al.}(2013){Cluver}, {Appleton}, {Ogle},
  {et~al.}}]{cluver+13}
{Cluver}, M.~E., {Appleton}, P.~N., {Ogle}, P., {et~al.} 2013, \apj, 765, 93

\bibitem[{{Combes} {et~al.}(2013){Combes}, {Garc{\'{\i}}a-Burillo}, {Braine},
  {et~al.}}]{combes+13}
{Combes}, F., {Garc{\'{\i}}a-Burillo}, S., {Braine}, J., {et~al.} 2013, \aap,
  550, A41

\bibitem[{{Combes} {et~al.}(2007){Combes}, {Young}, \& {Bureau}}]{combes+07}
{Combes}, F., {Young}, L.~M., \& {Bureau}, M. 2007, \mnras, 377, 1795

\bibitem[{{Condon}(1992)}]{condon92}
{Condon}, J.~J. 1992, \araa, 30, 575

\bibitem[{{Crocker} {et~al.}(2011){Crocker}, {Bureau}, {Young}, \&
  {Combes}}]{crocker+11}
{Crocker}, A.~F., {Bureau}, M., {Young}, L.~M., \& {Combes}, F. 2011, \mnras,
  410, 1197

\bibitem[{{Crocker} {et~al.}(2013){Crocker}, {Calzetti}, {Thilker},
  {et~al.}}]{crocker+13}
{Crocker}, A.~F., {Calzetti}, D., {Thilker}, D.~A., {et~al.} 2013, \apj, 762,
  79

\bibitem[{{da Cunha} {et~al.}(2008){da Cunha}, {Charlot}, \& {Elbaz}}]{magphys}
{da Cunha}, E., {Charlot}, S., \& {Elbaz}, D. 2008, \mnras, 388, 1595

\bibitem[{{Dale} {et~al.}(2012){Dale}, {Aniano}, {Engelbracht},
  {et~al.}}]{dale+12}
{Dale}, D.~A., {Aniano}, G., {Engelbracht}, C.~W., {et~al.} 2012, \apj, 745, 95

\bibitem[{{Davis} {et~al.}(2011){Davis}, {Alatalo}, {Sarzi},
  {et~al.}}]{davis+11}
{Davis}, T.~A., {Alatalo}, K., {Sarzi}, M., {et~al.} 2011, \mnras, 417, 882

\bibitem[{{Davis} {et~al.}(2012){Davis}, {Krajnovi{\'c}}, {McDermid},
  {et~al.}}]{davis+12}
{Davis}, T.~A., {Krajnovi{\'c}}, D., {McDermid}, R.~M., {et~al.} 2012, \mnras,
  426, 1574

\bibitem[{{Davis} {et~al.}(2014){Davis}, {Young}, {Crocker},
  {et~al.}}]{davis+14}
{Davis}, T.~A., {Young}, L.~M., {Crocker}, A.~F., {et~al.} 2014, \mnras, 444,
  3427

\bibitem[{{Deller} \& {Middelberg}(2014)}]{mjive}
{Deller}, A.~T., \& {Middelberg}, E. 2014, \aj, 147, 14

\bibitem[{{Dopita} {et~al.}(2002){Dopita}, {Pereira}, {Kewley}, \&
  {Capaccioli}}]{dopita+02}
{Dopita}, M.~A., {Pereira}, M., {Kewley}, L.~J., \& {Capaccioli}, M. 2002,
  \apjs, 143, 47

\bibitem[{{Downes} \& {Solomon}(1998)}]{downes+solomon98}
{Downes}, D., \& {Solomon}, P.~M. 1998, \apj, 507, 615

\bibitem[{{Draine} {et~al.}(2007){Draine}, {Dale}, {Bendo},
  {et~al.}}]{draine+07}
{Draine}, B.~T., {Dale}, D.~A., {Bendo}, G., {et~al.} 2007, \apj, 663, 866

\bibitem[{{Dressler} \& {Gunn}(1983)}]{dressler+gunn83}
{Dressler}, A., \& {Gunn}, J.~E. 1983, \apj, 270, 7

\bibitem[{{Dressler} {et~al.}(2013){Dressler}, {Oemler}, {Poggianti},
  {et~al.}}]{dressler+13}
{Dressler}, A., {Oemler}, Jr., A., {Poggianti}, B.~M., {et~al.} 2013, \apj,
  770, 62

\bibitem[{{Elvis} {et~al.}(1994){Elvis}, {Wilkes}, {McDowell},
  {et~al.}}]{elvis+94}
{Elvis}, M., {Wilkes}, B.~J., {McDowell}, J.~C., {et~al.} 1994, \apjs, 95, 1

\bibitem[{{Evans} {et~al.}(2005){Evans}, {Mazzarella}, {Surace},
  {et~al.}}]{evans+05}
{Evans}, A.~S., {Mazzarella}, J.~M., {Surace}, J.~A., {et~al.} 2005, \apjs,
  159, 197

\bibitem[{{Faber} {et~al.}(2007){Faber}, {Willmer}, {Wolf},
  {et~al.}}]{faber+07}
{Faber}, S.~M., {Willmer}, C.~N.~A., {Wolf}, C., {et~al.} 2007, \apj, 665, 265

\bibitem[{{Feruglio} {et~al.}(2015){Feruglio}, {Fiore}, {Carniani},
  {et~al.}}]{feruglio+15}
{Feruglio}, C., {Fiore}, F., {Carniani}, S., {et~al.} 2015, \aap, 583, A99

\bibitem[{{Feruglio} {et~al.}(2010){Feruglio}, {Maiolino}, {Piconcelli},
  {et~al.}}]{feruglio+10}
{Feruglio}, C., {Maiolino}, R., {Piconcelli}, E., {et~al.} 2010, \aap, 518,
  L155

\bibitem[{{Fischer} {et~al.}(2010){Fischer}, {Sturm}, {Gonz{\'a}lez-Alfonso},
  {et~al.}}]{fischer+10}
{Fischer}, J., {Sturm}, E., {Gonz{\'a}lez-Alfonso}, E., {et~al.} 2010, \aap,
  518, L41

\bibitem[{{French} {et~al.}(2015){French}, {Yang}, {Zabludoff},
  {et~al.}}]{french+15}
{French}, K.~D., {Yang}, Y., {Zabludoff}, A., {et~al.} 2015, \apj, 801, 1

\bibitem[{{Goto}(2005)}]{goto05}
{Goto}, T. 2005, \mnras, 357, 937

\bibitem[{{Goto}(2007)}]{goto07}
---. 2007, \mnras, 377, 1222

\bibitem[{{Guillard} {et~al.}(2015){Guillard}, {Boulanger}, {Lehnert},
  {et~al.}}]{guillard+15}
{Guillard}, P., {Boulanger}, F., {Lehnert}, M.~D., {et~al.} 2015, \aap, 574,
  A32

\bibitem[{{Harker} {et~al.}(2006){Harker}, {Schiavon}, {Weiner}, \&
  {Faber}}]{harker+06}
{Harker}, J.~J., {Schiavon}, R.~P., {Weiner}, B.~J., \& {Faber}, S.~M. 2006,
  \apjl, 647, L103

\bibitem[{{Hickson} {et~al.}(1992){Hickson}, {Mendes de Oliveira}, {Huchra}, \&
  {Palumbo}}]{hickson+92}
{Hickson}, P., {Mendes de Oliveira}, C., {Huchra}, J.~P., \& {Palumbo}, G.~G.
  1992, \apj, 399, 353

\bibitem[{{Holmberg}(1958)}]{holmberg58}
{Holmberg}, E. 1958, Meddelanden fran Lunds Astronomiska Observatorium Serie
  II, 136, 1

\bibitem[{{Hopkins} {et~al.}(2006){Hopkins}, {Hernquist}, {Cox},
  {et~al.}}]{hopkins+06}
{Hopkins}, P.~F., {Hernquist}, L., {Cox}, T.~J., {et~al.} 2006, \apjs, 163, 1

\bibitem[{{Hubble}(1926)}]{hubble26}
{Hubble}, E.~P. 1926, \apj, 64, 321

\bibitem[{{Ivezi{\'c}} {et~al.}(2002){Ivezi{\'c}}, {Menou}, {Knapp},
  {et~al.}}]{ivezic+02}
{Ivezi{\'c}}, {\v Z}., {Menou}, K., {Knapp}, G.~R., {et~al.} 2002, \aj, 124,
  2364

\bibitem[{{Jeong} {et~al.}(2013){Jeong}, {Yi}, {Kyeong}, {et~al.}}]{jeong+13}
{Jeong}, H., {Yi}, S.~K., {Kyeong}, J., {et~al.} 2013, \apjs, 208, 7

\bibitem[{{Kaneko} {et~al.}(2014){Kaneko}, {Kuno}, {Iono},
  {et~al.}}]{kaneko+15}
{Kaneko}, H., {Kuno}, N., {Iono}, D., {et~al.} 2014, \pasj\ submitted,
  arXiv:1411.2660

\bibitem[{{Kauffmann} {et~al.}(2003){Kauffmann}, {Heckman}, {Tremonti},
  {et~al.}}]{kauffmann+03}
{Kauffmann}, G., {Heckman}, T.~M., {Tremonti}, C., {et~al.} 2003, \mnras, 346,
  1055

\bibitem[{{Kenney} {et~al.}(2014){Kenney}, {Geha}, {J{\'a}chym},
  {et~al.}}]{kenney+14}
{Kenney}, J.~D.~P., {Geha}, M., {J{\'a}chym}, P., {et~al.} 2014, \apj, 780, 119

\bibitem[{{Kennicutt}(1998)}]{ken98}
{Kennicutt}, Jr., R.~C. 1998, \apj, 498, 541

\bibitem[{{Kewley} {et~al.}(2006){Kewley}, {Groves}, {Kauffmann}, \&
  {Heckman}}]{kewley+06}
{Kewley}, L.~J., {Groves}, B., {Kauffmann}, G., \& {Heckman}, T. 2006, \mnras,
  372, 961

\bibitem[{{Ko} {et~al.}(2013){Ko}, {Hwang}, {Lee}, \& {Sohn}}]{ko+13}
{Ko}, J., {Hwang}, H.~S., {Lee}, J.~C., \& {Sohn}, Y.-J. 2013, \apj, 767, 90

\bibitem[{{Kocevski} {et~al.}(2011){Kocevski}, {Lemaux}, {Lubin},
  {et~al.}}]{kocevski+11}
{Kocevski}, D.~D., {Lemaux}, B.~C., {Lubin}, L.~M., {et~al.} 2011, \apjl, 737,
  L38

\bibitem[{{Lanz} {et~al.}(2015){Lanz}, {Ogle}, {Alatalo}, \&
  {Appleton}}]{lanz+16}
{Lanz}, L., {Ogle}, P.~M., {Alatalo}, K., \& {Appleton}, P.~N. 2015, \apj\ in
  press, arXiv:1511.05968

\bibitem[{{Laor} \& {Behar}(2008)}]{laor+08}
{Laor}, A., \& {Behar}, E. 2008, \mnras, 390, 847

\bibitem[{{Lehnert} {et~al.}(2011){Lehnert}, {Tasse}, {Nesvadba}, {Best}, \&
  {van Driel}}]{lehnert+11}
{Lehnert}, M.~D., {Tasse}, C., {Nesvadba}, N.~P.~H., {Best}, P.~N., \& {van
  Driel}, W. 2011, \aap, 532, L3

\bibitem[{{Leon} {et~al.}(1998){Leon}, {Combes}, \& {Menon}}]{leon+98}
{Leon}, S., {Combes}, F., \& {Menon}, T.~K. 1998, \aap, 330, 37

\bibitem[{{Lisenfeld} {et~al.}(2014){Lisenfeld}, {Appleton}, {Cluver},
  {et~al.}}]{lisenfeld+14}
{Lisenfeld}, U., {Appleton}, P.~N., {Cluver}, M.~E., {et~al.} 2014, \aap, 570,
  A24

\bibitem[{{Lisenfeld} {et~al.}(2011){Lisenfeld}, {Espada}, {Verdes-Montenegro},
  {et~al.}}]{lisenfeld+11}
{Lisenfeld}, U., {Espada}, D., {Verdes-Montenegro}, L., {et~al.} 2011, \aap,
  534, A102

\bibitem[{{Martig} {et~al.}(2009){Martig}, {Bournaud}, {Teyssier}, \&
  {Dekel}}]{martig+09}
{Martig}, M., {Bournaud}, F., {Teyssier}, R., \& {Dekel}, A. 2009, \apj, 707,
  250

\bibitem[{{Martig} {et~al.}(2013){Martig}, {Crocker}, {Bournaud},
  {et~al.}}]{martig+13}
{Martig}, M., {Crocker}, A.~F., {Bournaud}, F., {et~al.} 2013, \mnras, 432,
  1914

\bibitem[{{Martin} {et~al.}(2007){Martin}, {Wyder}, {Schiminovich},
  {et~al.}}]{martin+07}
{Martin}, D.~C., {Wyder}, T.~K., {Schiminovich}, D., {et~al.} 2007, \apjs, 173,
  342

\bibitem[{{Martinez-Badenes} {et~al.}(2012){Martinez-Badenes}, {Lisenfeld},
  {Espada}, {et~al.}}]{martinez-badenes+12}
{Martinez-Badenes}, V., {Lisenfeld}, U., {Espada}, D., {et~al.} 2012, \aap,
  540, A96

\bibitem[{{McBride} {et~al.}(2014){McBride}, {Quataert}, {Heiles}, \&
  {Bauermeister}}]{mcbride+14}
{McBride}, J., {Quataert}, E., {Heiles}, C., \& {Bauermeister}, A. 2014, \apj,
  780, 182

\bibitem[{{Murray} {et~al.}(2007){Murray}, {Martin}, {Quataert}, \&
  {Thompson}}]{murray+07}
{Murray}, N., {Martin}, C.~L., {Quataert}, E., \& {Thompson}, T.~A. 2007, \apj,
  660, 211

\bibitem[{{Narayanan} {et~al.}(2011){Narayanan}, {Krumholz}, {Ostriker}, \&
  {Hernquist}}]{narayanan+11}
{Narayanan}, D., {Krumholz}, M., {Ostriker}, E.~C., \& {Hernquist}, L. 2011,
  \mnras, 418, 664

\bibitem[{{Nyland} {et~al.}(2013){Nyland}, {Alatalo}, {Wrobel},
  {et~al.}}]{nyland+13}
{Nyland}, K., {Alatalo}, K., {Wrobel}, J.~M., {et~al.} 2013, \apj, 779, 173

\bibitem[{{O'Connell}(1999)}]{oconnell99}
{O'Connell}, R.~W. 1999, \araa, 37, 603

\bibitem[{{Oh} {et~al.}(2011){Oh}, {Sarzi}, {Schawinski}, \& {Yi}}]{ossy}
{Oh}, K., {Sarzi}, M., {Schawinski}, K., \& {Yi}, S.~K. 2011, \apjs, 195, 13

\bibitem[{{Park} {et~al.}(2007){Park}, {Choi}, {Vogeley}, {et~al.}}]{park+07}
{Park}, C., {Choi}, Y.-Y., {Vogeley}, M.~S., {et~al.} 2007, \apj, 658, 898

\bibitem[{{Park} {et~al.}(2015){Park}, {Jeong}, \& {Yi}}]{park+15}
{Park}, J., {Jeong}, H., \& {Yi}, S.~K. 2015, \apj, 809, 91

\bibitem[{{Quintero} {et~al.}(2004){Quintero}, {Hogg}, {Blanton},
  {et~al.}}]{quintero+04}
{Quintero}, A.~D., {Hogg}, D.~W., {Blanton}, M.~R., {et~al.} 2004, \apj, 602,
  190

\bibitem[{{Rangwala} {et~al.}(2011){Rangwala}, {Maloney}, {Glenn},
  {et~al.}}]{rangwala+11}
{Rangwala}, N., {Maloney}, P.~R., {Glenn}, J., {et~al.} 2011, \apj, 743, 94

\bibitem[{{Rasmussen} {et~al.}(2008){Rasmussen}, {Ponman}, {Verdes-Montenegro},
  {Yun}, \& {Borthakur}}]{rasmussen+08}
{Rasmussen}, J., {Ponman}, T.~J., {Verdes-Montenegro}, L., {Yun}, M.~S., \&
  {Borthakur}, S. 2008, \mnras, 388, 1245

\bibitem[{{Rich} {et~al.}(2011){Rich}, {Kewley}, \& {Dopita}}]{rich+11}
{Rich}, J.~A., {Kewley}, L.~J., \& {Dopita}, M.~A. 2011, \apj, 734, 87

\bibitem[{{Rich} {et~al.}(2014){Rich}, {Kewley}, \& {Dopita}}]{rich+14}
---. 2014, \apjl, 781, L12

\bibitem[{{Rich} {et~al.}(2015){Rich}, {Kewley}, \& {Dopita}}]{rich+15}
---. 2015, \apjs, 221, 28

\bibitem[{{Rosenberg} {et~al.}(2015){Rosenberg}, {van der Werf}, {Aalto},
  {et~al.}}]{rosenberg+15}
{Rosenberg}, M.~J.~F., {van der Werf}, P.~P., {Aalto}, S., {et~al.} 2015, \apj,
  801, 72

\bibitem[{{Rowlands} {et~al.}(2015){Rowlands}, {Wild}, {Nesvadba},
  {et~al.}}]{rowlands+15}
{Rowlands}, K., {Wild}, V., {Nesvadba}, N., {et~al.} 2015, \mnras, 448, 258

\bibitem[{{Rupke} {et~al.}(2005){Rupke}, {Veilleux}, \& {Sanders}}]{rupke+05}
{Rupke}, D.~S., {Veilleux}, S., \& {Sanders}, D.~B. 2005, \apjs, 160, 115

\bibitem[{{Rupke} \& {Veilleux}(2011)}]{rupke+11}
{Rupke}, D.~S.~N., \& {Veilleux}, S. 2011, \apjl, 729, L27

\bibitem[{{Rupke} \& {Veilleux}(2015)}]{rupke+15}
---. 2015, \apj, 801, 126

\bibitem[{{Saintonge} {et~al.}(2011){Saintonge}, {Kauffmann}, {Kramer},
  {et~al.}}]{coldgass}
{Saintonge}, A., {Kauffmann}, G., {Kramer}, C., {et~al.} 2011, \mnras, 415, 32

\bibitem[{{Salom{\'e}} {et~al.}(2016){Salom{\'e}}, {Salom{\'e}}, {Combes},
  {Hamer}, \& {Heywood}}]{salome+16}
{Salom{\'e}}, Q., {Salom{\'e}}, P., {Combes}, F., {Hamer}, S., \& {Heywood}, I.
  2016, \aap, 586, A45

\bibitem[{{Sanders} {et~al.}(1989){Sanders}, {Phinney}, {Neugebauer}, {Soifer},
  \& {Matthews}}]{sanders+89}
{Sanders}, D.~B., {Phinney}, E.~S., {Neugebauer}, G., {Soifer}, B.~T., \&
  {Matthews}, K. 1989, \apj, 347, 29

\bibitem[{{Sandstrom} {et~al.}(2013){Sandstrom}, {Leroy}, {Walter},
  {et~al.}}]{sandstrom+13}
{Sandstrom}, K.~M., {Leroy}, A.~K., {Walter}, F., {et~al.} 2013, \apj, 777, 5

\bibitem[{{Sarzi} {et~al.}(2016){Sarzi}, {Kaviraj}, {Nedelchev},
  {et~al.}}]{sarzi+16}
{Sarzi}, M., {Kaviraj}, S., {Nedelchev}, B., {et~al.} 2016, \mnras, 456, L25

\bibitem[{{Sault} {et~al.}(1995){Sault}, {Teuben}, \& {Wright}}]{miriad}
{Sault}, R.~J., {Teuben}, P.~J., \& {Wright}, M.~C.~H. 1995, in Astronomical
  Society of the Pacific Conference Series, Vol.~77, Astronomical Data Analysis
  Software and Systems IV, ed. R.~A. {Shaw}, H.~E. {Payne}, \& J.~J.~E.
  {Hayes}, 433

\bibitem[{{Schawinski} {et~al.}(2014){Schawinski}, {Urry}, {Simmons},
  {et~al.}}]{schawinski+14}
{Schawinski}, K., {Urry}, C.~M., {Simmons}, B.~D., {et~al.} 2014, \mnras, 440,
  889

\bibitem[{{Scoville} {et~al.}(2014){Scoville}, {Aussel}, {Sheth},
  {et~al.}}]{scoville+14}
{Scoville}, N., {Aussel}, H., {Sheth}, K., {et~al.} 2014, \apj, 783, 84

\bibitem[{{Skrutskie} {et~al.}(2006){Skrutskie}, {Cutri}, {Stiening},
  {et~al.}}]{2mass}
{Skrutskie}, M.~F., {Cutri}, R.~M., {Stiening}, R., {et~al.} 2006, \aj, 131,
  1163

\bibitem[{{Solomon} \& {Vanden Bout}(2005)}]{solomon+vandenbout05}
{Solomon}, P.~M., \& {Vanden Bout}, P.~A. 2005, \araa, 43, 677

\bibitem[{{Spergel} {et~al.}(2007){Spergel}, {Bean}, {Dor{\'e}},
  {et~al.}}]{wmap}
{Spergel}, D.~N., {Bean}, R., {Dor{\'e}}, O., {et~al.} 2007, \apjs, 170, 377

\bibitem[{{Springel} {et~al.}(2005){Springel}, {Di Matteo}, \&
  {Hernquist}}]{springel+05}
{Springel}, V., {Di Matteo}, T., \& {Hernquist}, L. 2005, \apjl, 620, L79

\bibitem[{{Strateva} {et~al.}(2001){Strateva}, {Ivezi{\'c}}, {Knapp},
  {et~al.}}]{strateva+01}
{Strateva}, I., {Ivezi{\'c}}, {\v Z}., {Knapp}, G.~R., {et~al.} 2001, \aj, 122,
  1861

\bibitem[{{Sturm} {et~al.}(2011){Sturm}, {Gonz{\'a}lez-Alfonso}, {Veilleux},
  {et~al.}}]{sturm+11}
{Sturm}, E., {Gonz{\'a}lez-Alfonso}, E., {Veilleux}, S., {et~al.} 2011, \apjl,
  733, L16

\bibitem[{{Tinsley}(1978)}]{tinsley78}
{Tinsley}, B.~M. 1978, \apj, 222, 14

\bibitem[{{Toomre} \& {Toomre}(1972)}]{toomre72}
{Toomre}, A., \& {Toomre}, J. 1972, \apj, 178, 623

\bibitem[{{Vazdekis} {et~al.}(2010){Vazdekis}, {S{\'a}nchez-Bl{\'a}zquez},
  {Falc{\'o}n-Barroso}, {et~al.}}]{vazdekis+10}
{Vazdekis}, A., {S{\'a}nchez-Bl{\'a}zquez}, P., {Falc{\'o}n-Barroso}, J.,
  {et~al.} 2010, \mnras, 404, 1639

\bibitem[{{Veilleux} {et~al.}(2005){Veilleux}, {Cecil}, \&
  {Bland-Hawthorn}}]{veilleux+05}
{Veilleux}, S., {Cecil}, G., \& {Bland-Hawthorn}, J. 2005, \araa, 43, 769

\bibitem[{{Veilleux} \& {Osterbrock}(1987)}]{veilleux+87}
{Veilleux}, S., \& {Osterbrock}, D.~E. 1987, \apjs, 63, 295

\bibitem[{{Verdes-Montenegro} {et~al.}(1998){Verdes-Montenegro}, {Yun},
  {Perea}, {del Olmo}, \& {Ho}}]{verdes-montenegro+98}
{Verdes-Montenegro}, L., {Yun}, M.~S., {Perea}, J., {del Olmo}, A., \& {Ho},
  P.~T.~P. 1998, \apj, 497, 89

\bibitem[{{Ward} {et~al.}(1987){Ward}, {Elvis}, {Fabbiano}, {et~al.}}]{ward+87}
{Ward}, M., {Elvis}, M., {Fabbiano}, G., {et~al.} 1987, \apj, 315, 74

\bibitem[{{Wild} {et~al.}(2011){Wild}, {Charlot}, {Brinchmann},
  {et~al.}}]{wild+11}
{Wild}, V., {Charlot}, S., {Brinchmann}, J., {et~al.} 2011, \mnras, 417, 1760

\bibitem[{{Wild} {et~al.}(2009){Wild}, {Walcher}, {Johansson},
  {et~al.}}]{wild+09}
{Wild}, V., {Walcher}, C.~J., {Johansson}, P.~H., {et~al.} 2009, \mnras, 395,
  144

\bibitem[{{Wright} {et~al.}(2010){Wright}, {Eisenhardt}, {Mainzer},
  {et~al.}}]{wise}
{Wright}, E.~L., {Eisenhardt}, P.~R.~M., {Mainzer}, A.~K., {et~al.} 2010, \aj,
  140, 1868

\bibitem[{{Yan} {et~al.}(2006){Yan}, {Newman}, {Faber}, {et~al.}}]{yan+06}
{Yan}, R., {Newman}, J.~A., {Faber}, S.~M., {et~al.} 2006, \apj, 648, 281

\bibitem[{{Yao} {et~al.}(2003){Yao}, {Seaquist}, {Kuno}, \& {Dunne}}]{yao+03}
{Yao}, L., {Seaquist}, E.~R., {Kuno}, N., \& {Dunne}, L. 2003, \apj, 588, 771

\bibitem[{{Yesuf} {et~al.}(2014){Yesuf}, {Faber}, {Trump}, {et~al.}}]{yesuf+14}
{Yesuf}, H.~M., {Faber}, S.~M., {Trump}, J.~R., {et~al.} 2014, \apj, 792, 84

\bibitem[{{Young} {et~al.}(2011){Young}, {Bureau}, {Davis},
  {et~al.}}]{young+11}
{Young}, L.~M., {Bureau}, M., {Davis}, T.~A., {et~al.} 2011, \mnras, 414, 940

\bibitem[{{Young} {et~al.}(2014){Young}, {Scott}, {Serra}, {et~al.}}]{young+14}
{Young}, L.~M., {Scott}, N., {Serra}, P., {et~al.} 2014, \mnras, 444, 3408

\bibitem[{{Zabludoff} \& {Mulchaey}(1998)}]{zabludoff+mulchaey98}
{Zabludoff}, A.~I., \& {Mulchaey}, J.~S. 1998, \apj, 496, 39

\bibitem[{{Zabludoff} {et~al.}(1996){Zabludoff}, {Zaritsky}, {Lin},
  {et~al.}}]{zabludoff+96}
{Zabludoff}, A.~I., {Zaritsky}, D., {Lin}, H., {et~al.} 1996, \apj, 466, 104

\bibitem[{{Zensus}(1997)}]{zensus+97}
{Zensus}, J.~A. 1997, \araa, 35, 607

\bibitem[{{Zucker} {et~al.}(2016){Zucker}, {Walker}, {Johnson},
  {et~al.}}]{zucker+16}
{Zucker}, C., {Walker}, L.~M., {Johnson}, K., {et~al.} 2016, \apj, 821, 113

\end{thebibliography}


 \end{document}